\documentclass[useAMS,usenatbib]{mnras}

\usepackage{amsmath}
\usepackage{amssymb}
\usepackage{graphicx}
\usepackage{bm}
\usepackage{geometry}
\usepackage{natbib}

\def\nat{Nature\ }
\def\aap{Astron.\ Astrophys.\ }
\def\apj{Astrophys.\ J.\ }
\def\apjl{Astrophys.\ J.\ Lett.\ }
\def\apjs{Astrophys.\ J.\ Supp.\ }
\def\aj{Astron.\ J.\ }
\def\mnras{Mon.\ Not.\ Roy.\ Astron.\ Soc.\ }

\def\prd{Phys.\ Rev.\ D\ }

\def\apss{Astrophys.\ Space\ Sci.\ }

\def\jcap{J.\ Cosmol.\ Astropart.\ Phys.\ }

\begin{document}
\title[Discriminating electron sources by anisotropy]{Discriminating Local 
Sources of High-Energy Cosmic Electrons and Positrons by Current and Future 
Anisotropy Measurements}

\author[K. Fang et al.]{Kun Fang$^{1,2}$ , Xiao-Jun Bi$^{1,2}$, Peng-Fei 
Yin$^{1}$ \\
$^{1}$ Key Laboratory of Particle Astrophysics, Institute of High Energy 
Physics, Chinese Academy of Sciences, Beijing 100049, China\\
$^{2}$ School of Physical Sciences, University of Chinese Academy of Sciences, 
Beijing 100049, China\\
}

\maketitle

\begin{abstract}
The Fermi-LAT detects no significant anisotropy of the cosmic-ray (CR) 
electrons and positrons ($e^-+e^+$) with seven years of data, which provides 
the strongest restriction to the $e^-+e^+$ anisotropy up to now. As next 
generation CR observatory, HERD is expected to have a better capability of 
anisotropy detection than Fermi-LAT. In this paper, we discuss several models 
aimed to explain the AMS-02 data by the present and future anisotropy 
measurements. We find that the upper limits of Fermi-LAT disfavor Vela SNR as 
the dominant source in sub-TeV, while other cases that remain safe under the 
constraint of Fermi-LAT are expected to be distinguished from each other by 
HERD. We then discuss the possibilities of remarkable TeV spectral features, 
and test the corresponding anisotropies. We find the conditions under which 
the TeV model can have a prominent spectral feature and avoid the constraint 
of Fermi-LAT at the same time. Furthermore, the expected performance of HERD 
is sensitive enough to detect the anisotropies of all these TeV models, and 
even for the case of a featureless TeV spectrum. Thus HERD may play a crucial 
part in the study of the origin of cosmic electrons and positrons.
\end{abstract}

\section{Introduction}
\label{sec:intro}
The AMS-02 collaboration has respectively obtained the cosmic-ray (CR) 
electron and positron ($e^{\pm}$) spectra \citep{amselec}, owing to the strong 
capability of the sign-of-charge discrimination. These measurements provide 
stronger constraints on theoretical models of the cosmic lepton origin, 
compared with merely the measurement of the electron plus positron ($e^-+e^+$)
spectrum. The high-precision results of AMS-02 not only confirm the existence 
of the $e^\pm$ excess detected by the previous experiments 
\citep{2008Natur.456..362C,
2009Natur.458..607A}, but also indicate that the electron spectrum has a 
larger excess than that of the positron \citep{2014PhLB..728..250F, 
2015PhLB..749..267L, 2015PhRvD..91f3508L}. This extra excess of electrons can 
be interpreted as the spectral fluctuation brought by discrete local supernova 
remnants (SNRs) \citep{2015PhRvD..91f3508L}.

After the work of \cite{shen70}, many astrophysical models have been 
proposed, in which nearby SNRs are calculated separately from background SNRs 
\citep{1995PhRvD..52.3265A,koba04, mauro14}, in order to explain the CR 
observations. Along this approach, we have carefully investigated local CR 
sources and discussed their parameters of electron injection,
to fit all the leptonic data of AMS-02, and give predictions to $e^-+e^+$
spectrum beyond 1 TeV \citep[hereafter F17]{2017ApJ...836..172F}. 
Vela SNR is traditionally believed to be the most important local 
source around TeV, while other candidates also have the possibility 
to explain the extra electron excess. Although the possible candidates to 
explain the AMS-02 data are few \citep{2017ApJ...836..172F}, the current 
leptonic data of AMS-02 cannot discriminate among them.

However, things may be different when the anisotropy in the arrival direction 
of cosmic electrons is taken into account. Even two sources produce similar 
leptonic spectra, the angular distributions would be different due to their 
different positions. Also, the magnitude of the anisotropy of these sources 
may also be different. Therefore the anisotropy measurement can provide an 
unprecedentedly strong constraint on theoretical models. 
\citet{2013ApJ...772...18L} depicted the prospective of ascertaining the 
origin of the pulsar account for the positron excess, by the anisotropy 
measurement of atmospheric Cherenkov telescopes. More recently, 
\citet{2017JCAP...01..006M} performed a detailed analysis of local CR $e^\pm$ 
sources, and presented the corresponding anisotropies.

In March 2017, the Fermi-LAT collaboration published the latest 
result of anisotropy measurement with seven years of data 
\citep{2017PhRvL.118i1103A}. As no significant anisotropy has
been detected in any angular scale, Fermi-LAT provides the strongest
upper limits of anisotropy so far. Furthermore, the future Chinese Space 
Station based instrument HERD (the High Energy Cosmic-Radiation Detector) aims 
to measure the energy spectrum and anisotropy of $e^-+e^+$ up to 10 TeV 
\citep[HERD,][]{2014SPIE.9144E..0XZ}. HERD is planned to be launched before 
2025 with more than 10 years of lifetime and is expected to have a better 
sensitivity than Fermi-LAT.

In this work we discuss the possibility to discriminate the local 
sources of high energy $e^\pm$ by combining the leptonic spectra and 
the current or future anisotropy measurements. Our work has several important
differences with the previous studies, such as \citet{2017JCAP...01..006M}. 
First, we calculate the total anisotropies of theoretical models, including the 
anisotropy of the SNR background. The background component also contributes 
considerably to the total anisotropy and should not be omitted in the 
calculation. Second, when Vela cannot account for the electron excess, we adopt 
specific sources like Monogem Ring or Loop I as the dominant local source in 
sub-TeV to explain the AMS-02 data. Third, we also discuss possible spectral 
features generated by local sources above TeV, together with their 
anisotropies. Finally, we adopt not only the latest anisotropy limits given by 
Fermi-LAT, but also the expected HERD sensitivity in the model discussions.

This paper is organized as follows. In Section \ref{sec:method}, we review the 
$e^\pm$ sources in our framework and the propagation of $e^\pm$; then we 
introduce the calculation of the total anisotropies of theoretical models, and 
the expectation of the sensitivity of HERD. In Section \ref{sec:results}, we 
first discuss the models aimed to explain the AMS-02 data with the anisotropy 
limits of Fermi-LAT and the expected sensitivity of HERD; then we calculate the 
anisotropies corresponding to several models with remarkable TeV spectral 
features, and compare them with the anisotropy measurements. We give some 
discussions on other possibilities in Section \ref{sec:discussion}, and then 
conclude in the last section.

\section{Method}
\label{sec:method}

\subsection{Injection and Propagation of Galactic $e^\pm$}
In F17, we have already described our calculation of $e^{\pm}$ spectra at the 
Earth in detail, so in the following we just give a brief introduction to the 
injection spectrum of $e^\pm$ sources and the propagation of CR $e^\pm$.
In our framework, SNRs are the main contributors of CR electrons, while pulsar 
wind nebulae (PWNe) provide both high energy positrons and electrons. 
Secondary positrons are considered as the background of the positron spectrum.

For SNRs, the injection spectrum of electrons can be expressed as 
\begin{equation}
  Q(E)=Q_0{(E/{\rm 1\,GeV})}^{-\gamma}{\rm exp}(-E/E_c)\,,
 \label{eq:Q}
\end{equation}
where $Q_0$ is the normalization of the injection spectrum, $\gamma$ is the 
power-law spectral index, and $E_c$ is the cut-off energy. We divide SNRs into 
a group of known young and local sources and a background component with 
a continuous distribution. The background SNRs share a common injection 
spectrum. Their $Q_{\rm 0,bkg}$ and $\gamma_{\rm bkg}$ are determined by the 
fitting procedures in the next section, and $E_c$ is set to be 20 TeV. The 
local SNRs adopted in this work are shown in Table \ref{tab:SNRs}. Considering 
the age and distance of these individual SNRs, only three of them---Vela YZ, 
Loop I, and 
Monogem Ring---can potentially contribute to the electron excess in the sub-TeV 
region (F17). We leave the discussion of these three sources in the next 
section. For other SNRs, we constrain their parameters of the injection 
spectrum by multi-band electromagnetic observations, or just radio observations 
in the absence of X-ray and $\gamma$-ray measurements. One may refer to F17 for 
the detailed processes. 

\begin{table}
\centering
 \begin{tabular}{ccccc}
  \hline
  Name & $l$($^\circ$) & $b$($^\circ$) & $r$(kpc) & $t$(kyr) \\
  \hline
  G65.3+5.7 & 65.3 & $+5.7$ & 0.8 & 28 \\
  \hline
  Cygnus Loop & 74.0 & $-8.5$ & 0.54 & 10 \\
  \hline
  G114.3+0.3 & 114.3 & $+0.3$ & 0.7 & 7.7\\
  \hline
  R5 & 127.1 & $+0.5$ & 1.00 & 25\\
  \hline
  G156.2+5.7 & 156.2 & $+5.7$ & 1.00 & 20.5\\
  \hline
  HB9 & 160.9 & $+2.6$ & 0.8 & 5.5\\
  \hline
  Vela Jr. & 266.2 & $-1.2$ & 0.75 & 3\\
  \hline
  RX J1713.7-3946 & 347.3 & $-0.5$ & 1.00 & 1.6\\
  \hline
  Vela YZ & 263.9 & $-3.3$ & 0.29 & 11.3\\
  \hline
  Monogem Ring & 203.0 & $+12.0$ & 0.3 & 86\\
  \hline
  Loop I (NPS) & 328.3 & $+17.6$ & 0.1 & 200\\
  \hline
 \end{tabular}
 \caption{The name, location, distance, and age of SNRs within 1 kpc. One can
refer to \citet{mauro14} and references therein for parameters of these sources;
while for Monogem Ring and Loop I, their distance and age listed here are taken
from \citet{plucinsky09} and \citet{egger95} respectively. }
\label{tab:SNRs}
\end{table}

The injection spectrum for the PWN case takes the same form with equation 
(\ref{eq:Q}), while PWNe can provide both electrons and positrons. As in F17, 
we adopt a single powerful PWN to explain the high energy positron spectrum of 
AMS-02. The only difference is that we choose Geminga as the single PWN in the 
present work, rather than Monogem. The characteristic age, distance, and 
spin-down luminosity of Geminga pulsar are $t=342$ kyr, $r=250$ pc, and 
$\dot{E}=3.25\times10^{34}$ erg s$^{-1}$, respectively 
\citep{2005AJ....129.1993M}. The total spin-down luminosity can be derived by 
$W_{\rm pwn}=\dot{E}t(1+t/\tau_0)=1.23\times10^{49}$ erg, where the spin-down 
time scale $\tau_0$ is set to be 10 kyr.  The normalization of the injection 
spectrum is decided by $W_{\rm pwn}$ and $\eta_{\rm pwn}$, the latter of which 
is the efficiency of energy conversion to the injected electrons and positrons. 
We set $\eta_{\rm pwn}$ and the spectral index $\gamma_{\rm pwn}$ as free 
parameters in the following fitting procedures. The calculation of secondary 
positrons keeps the same with that in F17.


The propagation of Galactic $e^\pm$ can be described by the diffusion equation 
with consideration of energy loss during their journey:
\begin{equation}
 \frac{\partial N}{\partial t} - \nabla(D\nabla N) - \frac{\partial}{\partial 
E}(bN) = Q \,,
 \label{eq:diff}
\end{equation}
where $N$ is the number density of $e^\pm$, $D$ denotes the diffusion 
coefficient, $b$ denotes the energy-loss rate, and $Q$ is the source function.
The diffusion coefficient $D$ has the form of $D(E)=\beta D_0{(R/\rm 
1\,GV)}^{\delta}$, where $D_0$ and $\delta$ are both constants, $\beta$ is the 
velocity of particles in the unit of light speed and $R$ is the rigidity of 
CRs. In this work, the diffusion coefficient refers to the DR2 model of 
\citet{2017PhRvD..95h3007Y}, in which $D_0=(2.08\pm0.28)\times10^{28}$
cm$^2$ s$^{-1}$, $\delta=0.500\pm0.012$. The energy-loss rate is given by 
$b(E)=-b_0(E)E^2$, where $b_0(E)$ is decided by synchrotron and inverse Compton 
radiation of CR $e^{\pm}$. We follow the method of \citet{schli10} to calculate 
the inverse Compton term, while we set the interstellar magnetic field in the 
Galaxy to be $1~\mu$G to get the synchrotron term \citep{1994A&A...288..759H, 
dela10}. 

Equation (\ref{eq:diff}) can be solved with the method of Green's function 
\citep{1964ocr..book.....G}. We adopt the spherically symmetric solution, which 
is safe for the case of high energy $e^{\pm}$ \citep{koba04}. To determine 
the source function, local SNRs and PWNe are considered as point sources with 
burst-like injection. For the background SNRs, we assume a smooth spatial 
distribution derived by \citet{l04}, and the SN explosion rate is set to be 
$f=4$ century$^{-1}$ galaxy$^{-1}$ \citep{dela10}. The critical distance and 
age between local and background SNRs are set to be $r_m=1$ kpc and 
$t_m=3\times10^5$ years respectively as what we do in F17.

\subsection{Anisotropy of Electrons and Positrons}
Presupposing the dipolar distribution of the intensity of CRs, the anisotropy
of CRs is generally defined as
\begin{equation}
 \Delta=\frac{I_{\rm max}-I_{\rm min}}{I_{\rm max}+I_{\rm min}}\,,
 \label{eq:ani}
\end{equation}
where $I_{\rm max}$ and $I_{\rm min}$ are the maximum and minimum values of the 
CR intensity at the Earth, respectively. If we take a specific form of the 
angular distribution of the CR intensity from a source as
\begin{equation}
I_i(\bm{n})=\bar{I_i}(1+\Delta_i\,\bm{n}_i\cdot\bm{n})\,,
\label{eq:intens_i}
\end{equation}
where $\bar{I_i}=(I_{i,{\rm max}}+I_{i,{\rm min}})/2$, ${\bm{n}}_i$ is the 
direction of the source, and $\Delta_i$ denotes the anisotropy of that source, 
then under the diffusion model, equation (\ref{eq:ani}) can be rewritten as
\begin{equation}
 \Delta_i=\frac{3D}{c}\cdot\frac{|\nabla N_i|}{N_i}\,,
 \label{eq:ani_s}
\end{equation}
which is derived by \citet{1964ocr..book.....G}. Combining equation 
(\ref{eq:ani_s}) and the solution of a point source with burst-like 
injection, we get the explicit expression of the $e^{\pm}$ anisotropy of a 
single source with distance $r_i$ and age $t_i$:
\begin{equation}
 \Delta_i=\frac{3D}{c}\cdot\frac{2r_i}{\lambda^2_i}\,,
 \label{eq:ani_s2}
\end{equation}
where 
\begin{equation}
\lambda_i^2=4\int_{E}^{\frac{E}{1-b_0Et_i}}\frac{D(E')}{b(E')}dE'
 \label{eq:lambda}
\end{equation}
describes the propagation distance of $e^{\pm}$ with arrival energy $E$.
If $E\ll 1/(b_0t_i)$, the diffusion scale can be approximated by
$\lambda_i=2\sqrt{D\,t_i}$. For a source with age of 10 kyr, this approximation
only brings a relative error of several percent to the anisotropy even in TeV
range. Under this approximation, equation (\ref{eq:ani_s2}) has a form of
\begin{equation}
 \Delta_i=\frac{3r_i}{2ct_i}\,,
 \label{eq:ani_s3}
\end{equation}
in which the anisotropy is simply decided by the age and distance of the 
source. We still adopt equation (\ref{eq:ani_s2}) to calculate the anisotropy 
in the following, but we should keep in mind that the anisotropy only has a 
weak dependence on the diffusion coefficient $D$.

In order to comparing theoretical models with experimental data, it is essential
to calculate the total anisotropy contributed by all the sources in each model.
The total intensity is given by summation of equation (\ref{eq:intens_i}):
\begin{equation}
I(\bm{n})=\sum_i\bar{I_i}(1+\Delta_i\bm{n}_i\cdot\bm{n})\,.
\label{eq:intens_t}
\end{equation}
Then the total anisotropy can be obtained from the definition of equation 
(\ref{eq:ani}):
\begin{equation}
\Delta=\frac{\sum\bar{I_i}\Delta_i\bm{n}_i\cdot\bm{n}_{\rm 
max}}{\sum\bar{I_i}}\, ,
\label{eq:ani_t}
\end{equation}
where $\bm{n}_{\rm max}$ is the direction of the maximum intensity. For 
the local SNRs, their locations can be found in the Green catalog. As to 
pulsars, their Galactic longitudes $l$ and latitudes $b$ are given in the ATNF 
catalog. Equation (\ref{eq:ani_t}) indicates that the contribution of an 
individual source to the total anisotropy is related to its relative intensity 
compared with the total intensity, rather than its absolute intensity. 

Since the distribution of distant SNRs is symmetric around the Galactic
center but not the solar system, and they also have significant contribution to
the electron spectrum, we emphasize that the SNR background should be a
considerable component in the calculation of the total anisotropy. We treat the 
SNR background component as a whole, so its anisotropy may also be calculated 
by equation (\ref{eq:ani_s}):
\begin{equation}
\Delta_{\rm 
bkg}=\frac{3D}{c}\cdot\frac{1}{N_{\rm bkg}}\cdot\left|\frac{N_{\rm 
bkg}(r_\odot+\Delta r)-N_{\rm bkg}(r_\odot-\Delta r)}{2\Delta r}\right|\,,
\label{eq:ani_bkg}
\end{equation}
where $N_{\rm bkg}$ is the number density of the background SNR. The difference 
step $\Delta r$ is set to be 0.1 kpc, and a smaller $\Delta r$ makes little 
difference to the result. The location of the background component is $l=0$, 
$b=0$, and hence the direction vector in the rectangular coordinate is 
$\bm{n}_{\rm bkg}=\{1, 0, 0\}$.

For each model in the following, we need to find $\bm{n}_{\rm max}$ in 
different energies. We calculate $I(l,b)$ for every $1^\circ\times1^\circ$ 
grid to find the maximum $I(l,b)$ and thus obtain $\bm{n}_{\rm max}$. The 
positions of all our local SNRs are listed in Table \ref{tab:SNRs}, while the 
location of PWNe discussed in the following sections can be found in Table 
\ref{tab:PWNe}.

\subsection{Expected Sensitivity of HERD}
HERD is a next generation CR observatory onboard China's Space 
Station, and is planned for operation starting around 2025 for about 10 
years. One of the main scientific objectives of HERD is the precise and direct 
measurements of the $e^-+e^+$ spectrum and the anisotropy up to 10 TeV. Here we 
give an expectation on the sensitivity of $e^-+e^+$ detection of HERD, and we 
will show in the following sections that the anisotropy measurement of HERD can 
play an important role in the study of CR $e^\pm$ origin. 

We briefly summarize the baseline performance of HERD first; one can refer to
\citet{2014SPIE.9144E..0XZ,herd1,herd2} for details. The main instrument of 
HERD is a 3-D cubic calorimeter (CALO). It is composed of thousands of small 
cubic crystals, and is surrounded by microstrip silicon trackers from five 
sides except the bottom. This novel design brings HERD an effective geometrical 
factor of $>3$ m$^2$ sr, which is 10 times larger than the previous CR 
detectors in space. The large geometrical factor is the biggest advantage of 
HERD, and it is the key parameter for the detection of anisotropy. The 
geometrical acceptances at different energies are shown in Table 
\ref{tab:herd}\footnote{These are the preliminary result provided by the HERD 
collaboration.}. Besides, by using 3-d imaging of the shower of the particle 
event, the average electron/proton (e/p) discrimination between 100 GeV and 1 
TeV is expected to reach $10^6$, which is much better than that of Fermi-LAT. 
HERD will measure the $e^-+e^+$ spectrum from 100 MeV to 10 TeV, with an energy 
resolution of 1\% at 100 GeV.

\begin{table}
\centering
 \begin{tabular}{cccccc}
  \hline
  Energy [GeV] & 100 & 200 & 500 & 750 & 1000 \\
  \hline
  A [m$^2$ sr] & 3.38 & 3.28 & 3.18 & 3.13 & 3.10 \\
  \hline
 \end{tabular}
 \caption{Geometrical acceptance of HERD for $e^\pm$ at different energies.}
\label{tab:herd}
\end{table}

We then estimate the sensitivity of $e^-+e^+$ detection of HERD. The detected 
number of $e^-+e^+$ can be calculated by $N=A\,t\,\bar{I}(E)\,\Delta E$, where 
$A$ is the energy dependent geometrical acceptance, which is obtained by 
interpolation or extrapolation with the data in Table \ref{tab:herd}. 
The operation time is denoted with $t$, and $\Delta E$ is the width of energy 
bins taken to be 0.3 in logarithmic scale. The flux of $e^-+e^+$ is taken as 
the theoretical $e^-+e^+$ spectrum in each theoretical model. The relative 
systematic uncertainty is simply $1/\sqrt{N}$, while the systematic uncertainty 
mainly arises from the misidentification of other particles as $e^\pm$ 
\citep{2010JCAP...12..020P}. We calculate the relative systematic uncertainty 
by $(N_p/N_e)/(e/p)$, where $N_p$ is the proton flux, $N_e$ is the flux of 
$e^-+e^+$, and $e/p$ is $10^6$ as described above. We adopt the proton flux 
measured by AMS-02 \citep{2015PhRvL.114q1103A}, and find the relative 
systematic uncertainty is in $\sim10^{-4}$. Thus the error is dominated by the 
statistical one, and the minimum detectable anisotropy at $2\sigma$ can be 
estimated by $2/\sqrt{N}$.

\section{Results}
\label{sec:results}
In F17 we have proposed several models to fit the AMS-02 data (hereafter
referred as the sub-TeV region). In Section \ref{subsec:subteV}, we 
emphasize on the corresponding anisotropy predictions of these models and show 
how the present and future measurements of anisotropy can constrain or 
discriminate among these models. Then the possibilities of prominent spectral 
features above TeV (hereafter referred as the TeV region) are discussed in 
Section \ref{subsec:TeV}, and the corresponding anisotropies are tested.


\subsection{Sub-TeV Region}
\label{subsec:subteV}

In the following, Vela YZ, Loop I, and Monogem Ring are taken as the 
predominant SNR in sub-TeV respectively. As described in the previous 
section, some of the parameters are decided by the global fitting to the four 
groups of leptonic data of AMS-02. However, for the sake of brevity, only the 
$e^-+e^+$ spectrum of each model is shown in the following figures.

\begin{table*}
 \raggedright
\begin{tabular}{ccccccc}
  \hline
  Model & $\gamma_{\rm bkg}$ & $Q_{0,{\rm bkg}}[10^{50}{\rm GeV^{-1}}]$ &  
$\gamma_{\rm
pwn}$ & $\eta_{\rm pwn}$ & $\gamma_{\rm vela}$ & $\chi^2/{\rm 
d.o.f}$\\
  \hline
  Vela & 2.42 & 2.48 & 1.82 & 1.80 & 2.12 & 0.43\\
  \hline
\end{tabular}
\begin{tabular}{ccccccccc}
  \hline
  Model & $\gamma_{\rm bkg}$ & $Q_{0,{\rm bkg}}[10^{50}{\rm GeV^{-1}}]$ &  
$\gamma_{\rm
pwn}$ & $\eta_{\rm pwn}$ & $\gamma_{\rm mr}$ & $W_{\rm
mr}$[$10^{48}\,$erg] & $E_{c,\rm mr}$[TeV] & $\chi^2/{\rm d.o.f}$ \\
  \hline
  Loop I & 2.50 & 3.11 & 1.79 & 1.72 & 1.87 & 6.29 & 1.55 &
0.42\\
  \hline
\end{tabular}
\begin{tabular}{ccccccccc}
  \hline
  Model & $\gamma_{\rm bkg}$ & $Q_{0,{\rm bkg}}[10^{50}{\rm GeV^{-1}}]$ & 
$\gamma_{\rm
pwn}$ & $\eta_{\rm pwn}$ & $\gamma_{\rm loop}$ & $W_{\rm
loop}$[$10^{48}\,$erg] & $E_{c,\rm loop}$[TeV] & $\chi^2/{\rm d.o.f}$ \\
  \hline
  MR & 2.51 & 3.35 & 1.79 & 1.70 & 1.93 & 2.24 & 1.09 &
0.42\\
  \hline
\end{tabular}
\caption{Fitting results of the sub-TeV models. Vela YZ is abbreviated to Vela 
in this table.}
\label{tab:sub-TeV}
\end{table*}

\begin{figure*}
\centering
\includegraphics[width=0.4\textwidth]{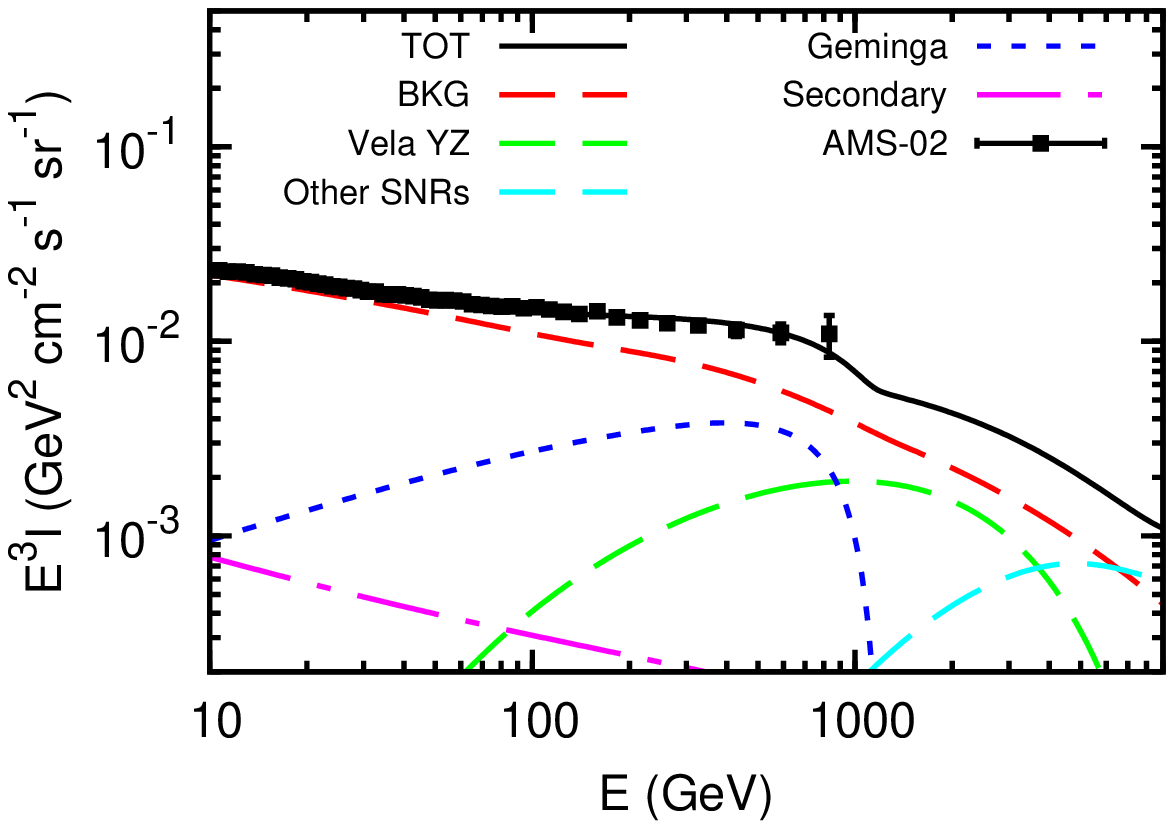}
\includegraphics[width=0.4\textwidth]{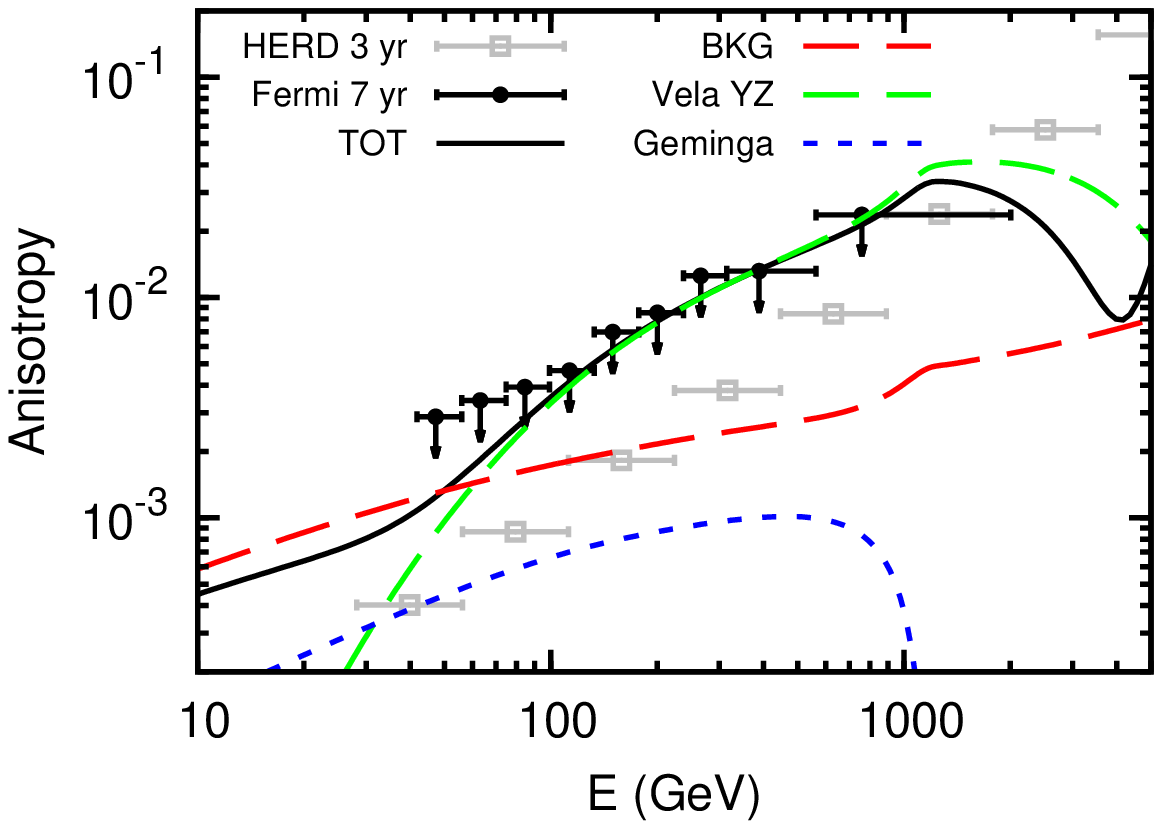}
\includegraphics[width=0.55\textwidth]{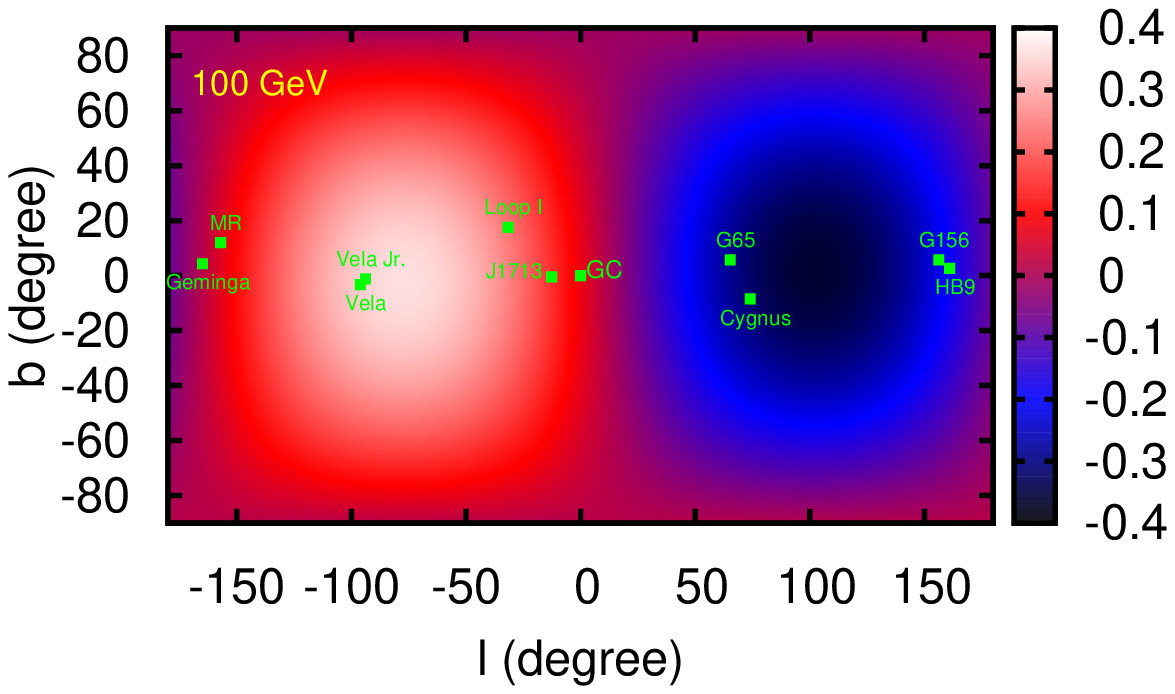}
\caption{The $e^-+e^+$ spectrum (top left), anisotropy (top right), and the 
angular distribution of intensity ($[I(l,b)-\bar{I}]/\bar{I}\times100$) at 100 
GeV (bottom) of $\textit{Vela YZ}$ model. In the legend, 'Other SNRs' refers to 
the summation of our local SNRs except for Vela YZ, Monogem Ring, and Loop I. 
In the top-right graph, the dash and dotted lines represent for the quantity 
$\bar{I_i}\Delta_i/\sum\bar{I_i}$. Note $\bar{I_i}\Delta_i/\sum\bar{I_i}$ of 
'Other SNRs' are not shown in the bottom graph, but they are included in the 
calculation of the total anisotropy. For the upper limits of Fermi-LAT, we 
adopt the result using the log-likelihood ratio method.}
\label{fig:vela}
\end{figure*}

\subsubsection{Vela YZ}
Vela SNR is believed to be the most promising local electron source, since its 
appropriate age and distance and its strong radio flux may lead to a 
significant 
contribution to the electron spectrum \citep{koba04,mauro14}. Vela SNR mainly 
consists of a PWN---Vela X, and two shell structure---Vela Y and Vela Z 
\citep{rish58}. We do not include Vela X here since it cannot help to explain 
the extra excess of electrons compared with positrons, due to its PWN nature 
\citep{weiler80}. In this model, we take some typical parameters for Vela 
YZ: a magnetic field of 30 $\mu$G, a cut-off energy of 2 TeV, and a radio flux 
of 1000 Jy at 1 GHz. The spectral index $\gamma_{\rm vela}$ is left to be free 
as it is a crucial parameter to the dominance of Vela YZ but still with large 
uncertainty in observation.  We seek the best-fit model by minimizing 
chi-square statistic between model and AMS-02 leptonic spectra\footnote{Steven 
G. Johnson, The NLopt nonlinear-optimization package, 
http://ab-initio.mit.edu/nlopt}. The best-fit parameters of this subsection are 
all listed in Table \ref{tab:sub-TeV}. 

Figure \ref{fig:vela} shows the best-fit $e^-+e^+$ spectrum, the corresponding 
anisotropy, and the angular distribution of intensity (to be exact, 
$[I(l,b)-\bar{I}]/\bar{I}\times100$) at 100 GeV. As can be seen, although 
the flux of Vela YZ is smaller than the background component, it still produces 
such a remarkable anisotropy that the total theoretical anisotropy reaches the 
exclusion limit at 95\% C.L. indicated by Fermi-LAT. Vela YZ leads the 
anisotropy from around 100 GeV to 1 TeV, above which its anisotropy is somehow 
offset by that of G65.3+5.7 whose position is opposite to Vela YZ. Vela YZ 
dominates the 
anisotropy not only because its considerable flux of electrons, but also its 
relatively large $r/t$ ratio, as explained by equation (\ref{eq:ani_s3}). The 
result indicates that the upper limits given by Fermi-LAT disfavor Vela YZ as 
the dominant electron source in sub-TeV. More generally, the extra electron 
excess in sub-TeV should not be explained by a relatively young SNR like Vela.
We also find that the expected sensitivity of three years of HERD data is 
obviously lower than the theoretical anisotropy of this model, which can give 
a more convincing judgment to this scenario in the future.

\begin{figure*}
\centering
\includegraphics[width=0.4\textwidth]{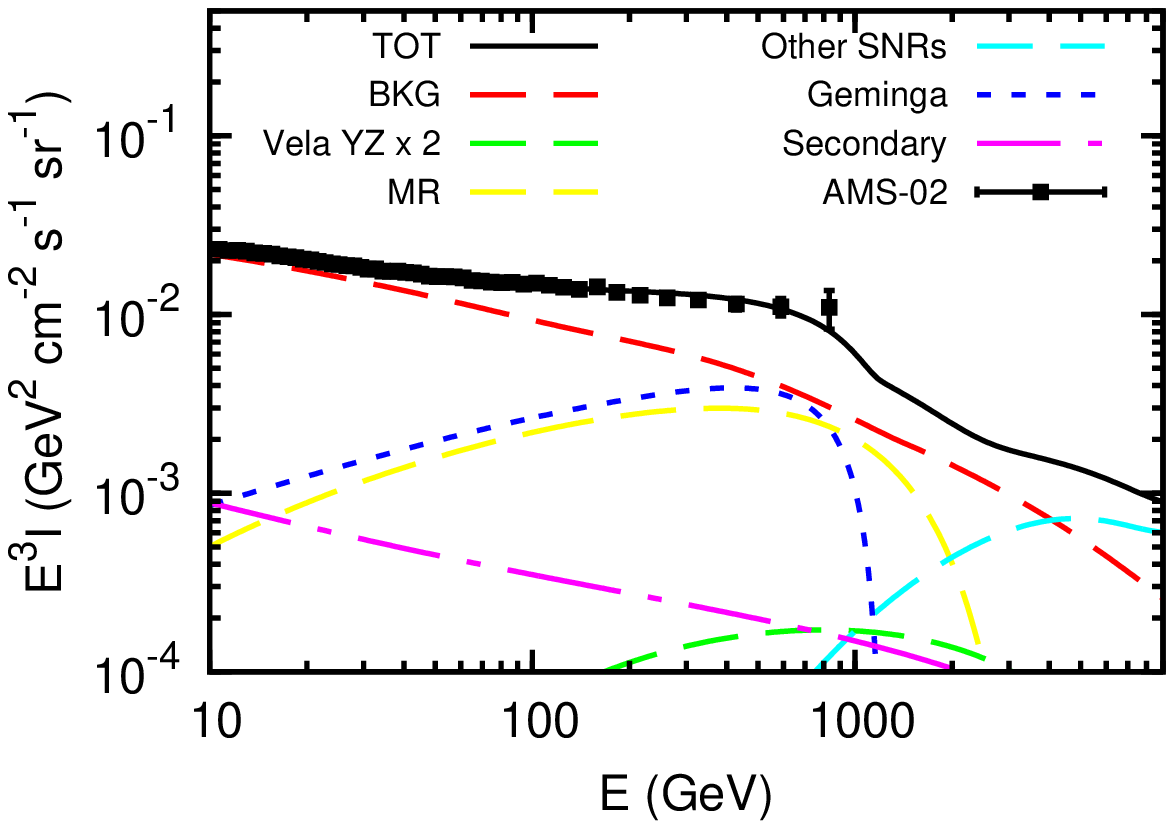}
\includegraphics[width=0.4\textwidth]{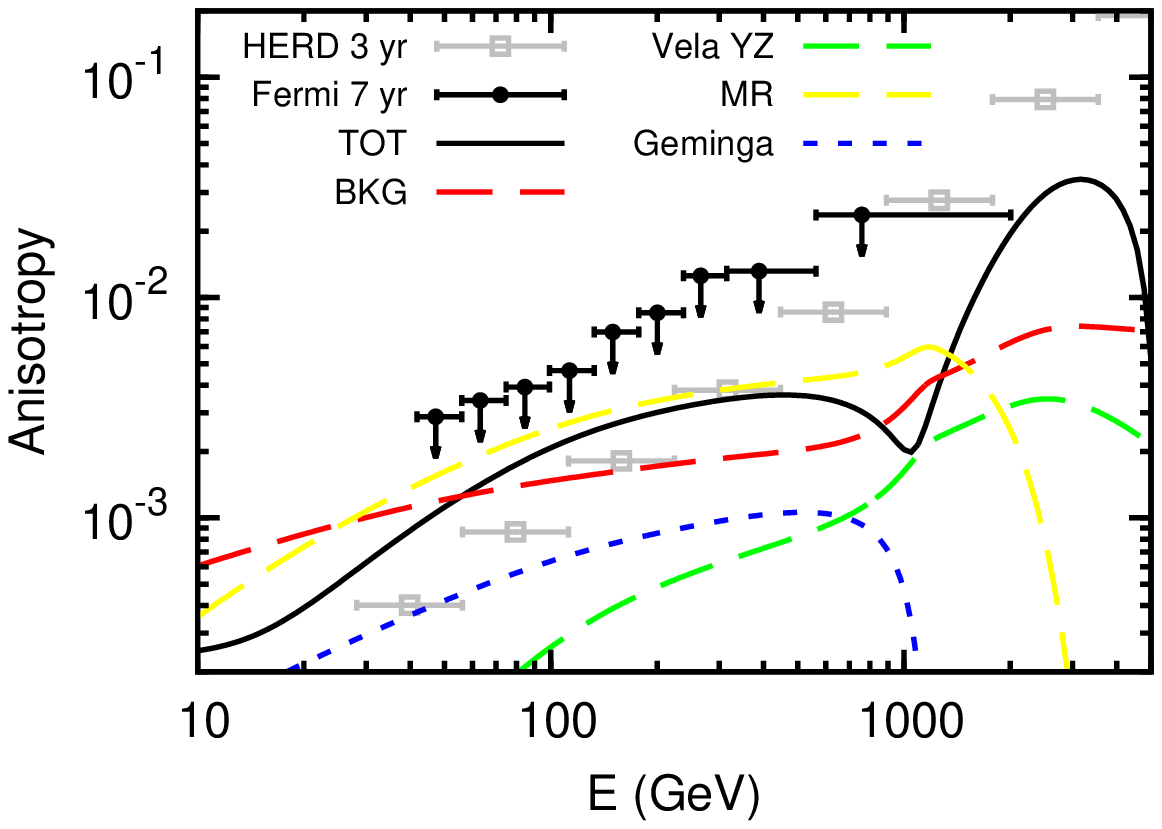}
\includegraphics[width=0.55\textwidth]{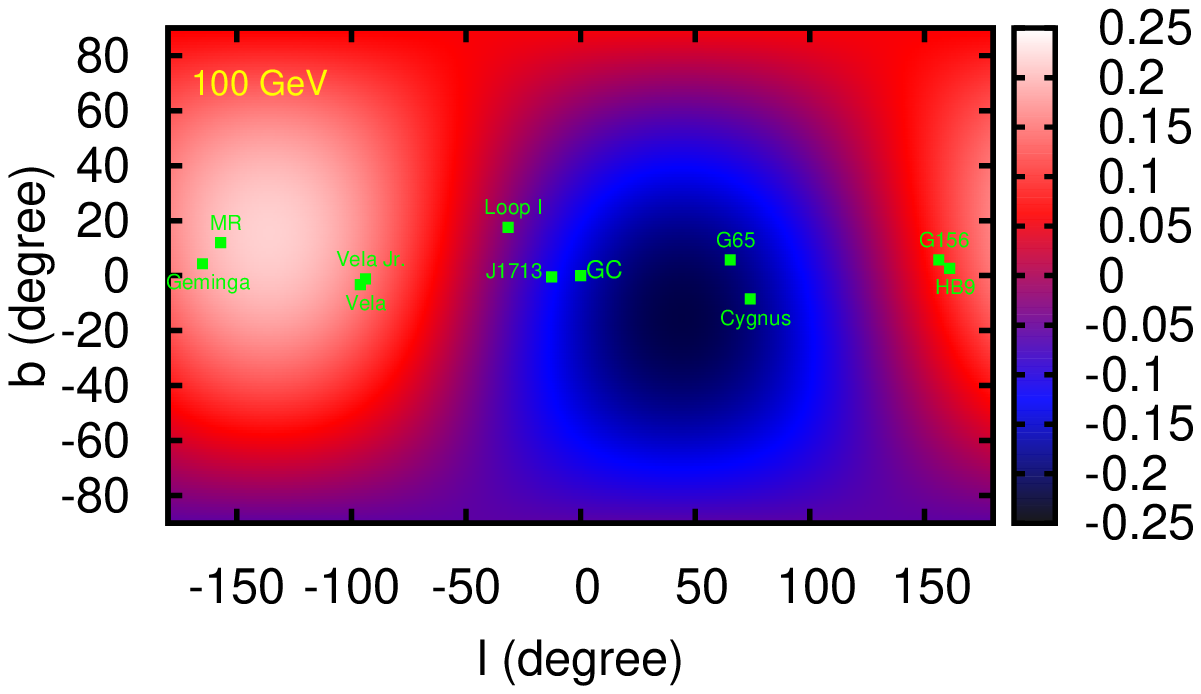}
\caption{Same as Figure \ref{fig:vela}, but for the \textit{Monogem Ring} 
scenario.}
\label{fig:mr}
\end{figure*}

\subsubsection{Monogem Ring}
As we introduced in F17, the electron flux of Vela YZ can be much smaller if we 
investigate a little more on its parameters. \citet{sushch14} derive a 
magnetic field of 46 $\mu$G in the region where Vela YZ is located. A shock 
velocity of $6\times10^7$ cm s$^{-1}$ is suggested by \citet{sushch11}, which 
leads to $E_c=4$ TeV \citep{yamazaki06}. The latest result of the radio 
spectrum of Vela Y and Vela Z can be found in \citet{alvarez01}. 
\citet{sushch14} merge the radio spectrum of Vela Y and Vela Z and obtain a 
electron spectral index of 2.47. In this case, the electron flux of Vela YZ is 
about 10 times less than the previous model, mainly due to the much softer 
spectral index. To explain the extra excess of electrons, Monogem Ring (MR) is 
one of the two candidate local SNRs.

MR is considered as the counterpart of pulsar B0656+14 which is located 
288 pc away \citep{plucinsky09}. A distance of $\sim300$ pc corresponds to an 
age of $8.6\times10^4$ years and an initial explosion energy of 
$0.19\times10^{51}$ erg, which is estimated by the X-ray observation and the 
Sedov-Taylor model \citep{plucinsky96}. We let the spectral index, total
energy, and cut-off energy free in the fitting, and denote them with 
$\gamma_{\rm mr}$, $W_{\rm mr}$, and $E_{c, {\rm mr}}$. As can be seen 
in Table \ref{tab:sub-TeV}, the best-fit electron energy is $2.24\times10^{48}$ 
erg. Relating this $W_{\rm mr}$ and the initial explosion energy of
$0.19\times10^{51}$ erg, a very large conversion efficiency of $10^{-2}$
is required.

The total anisotropy of this model is smaller than the sensitivity of 
Fermi-LAT, as shown in Fig. \ref{fig:mr}. This indicates that the 
\textit{Monogem Ring} scenario is still safe under the constraint of Fermi-LAT. 
As can be seen from the top right graph, MR contributes most to the total 
anisotropy in the sub-TeV region. At 100 GeV, the $\bm{n}_{\rm max}$ is near 
the direction of Monogem Ring, which can be found in the bottom graph of Fig. 
\ref{fig:mr}. Moreover, the theoretical anisotropy below 200 GeV of this model 
is larger than the expected sensitivity of HERD with 3 years observation, which 
means this model may also be tested by HERD.

\begin{figure*}
\centering
\includegraphics[width=0.4\textwidth]{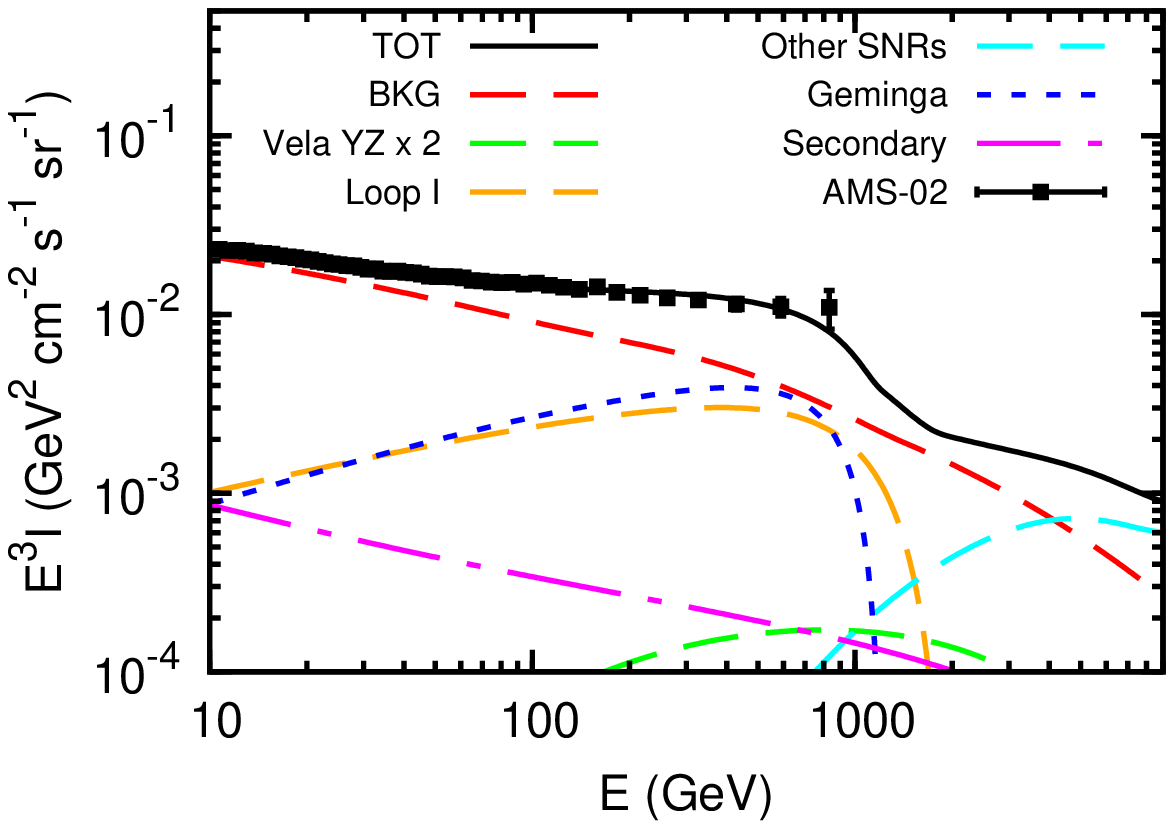}
\includegraphics[width=0.4\textwidth]{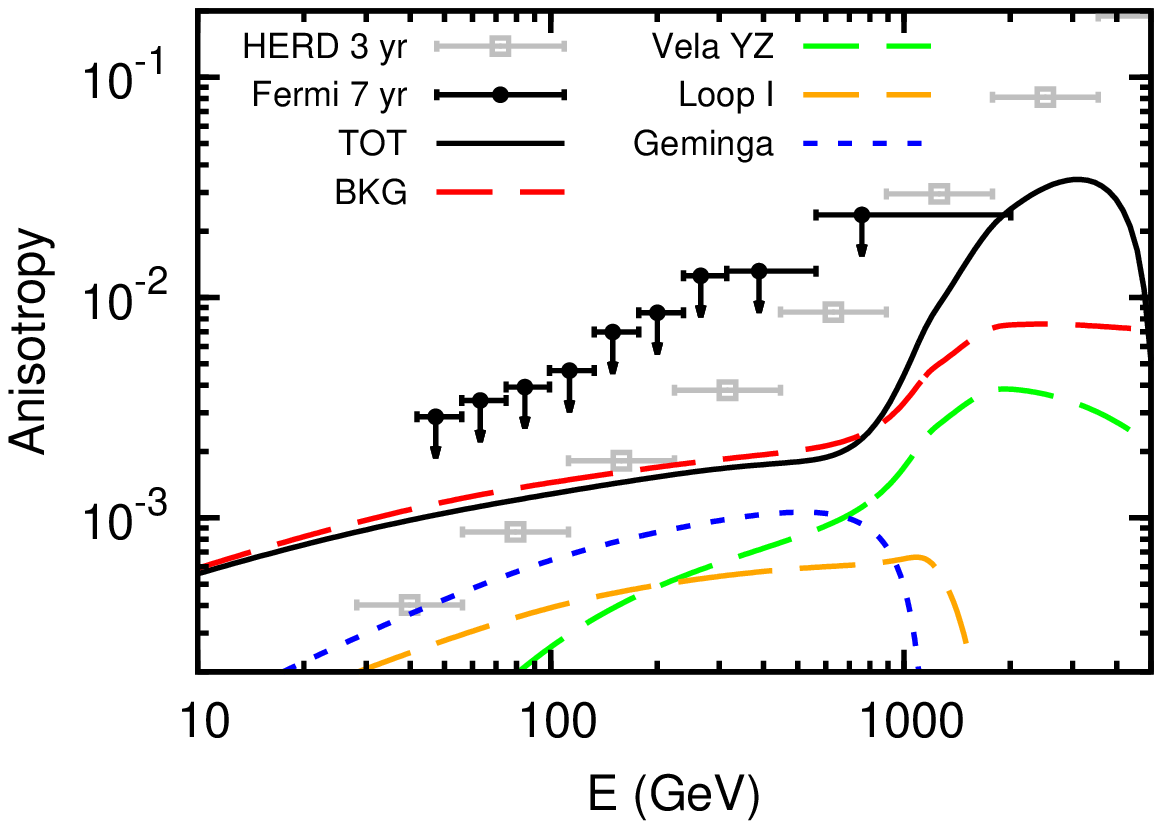}
\includegraphics[width=0.55\textwidth]{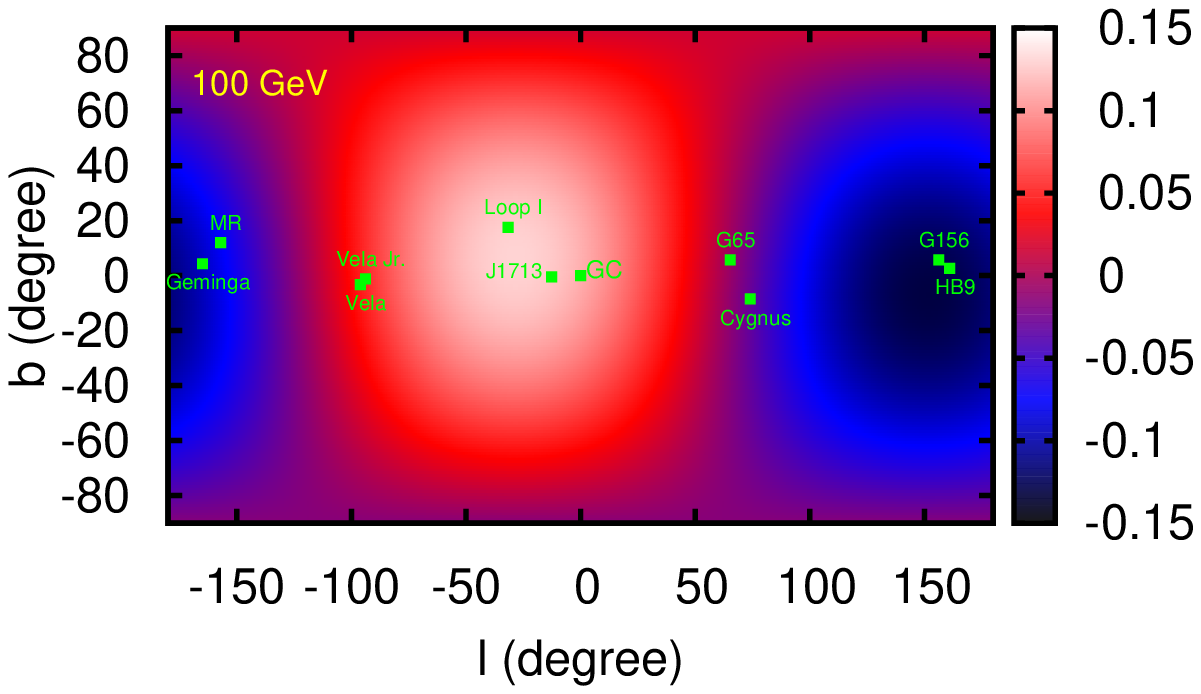}
\caption{Same as Figure \ref{fig:vela}, but for the \textit{Loop I (NPS)} 
scenario.}
\label{fig:loop1}
\end{figure*}

\subsubsection{Loop I (NPS)}
In addition to MR, Loop I is the other potential electron contributor in 
sub-TeV, when Vela YZ provides little flux of electrons. Although Loop I is 
believed to be an old structure ($\sim 10^6$ years) due to
the low velocity of neutral gas surrounding it \citep{sofue74}, the soft X-ray 
emission from the interior of Loop I indicates that there may be one or more 
subsequent SN events in the Sco-Cen association \citep{borken77, egger95}, 
which is located in the center of Loop I . The North Polar Spur (NPS) is the 
most prominent structure of Loop I, both in radio and X-ray maps, which can be 
interpreted by the reheating of a recent SN \citep{egger95}. What
is more, Fermi-LAT collaboration also detect high energy $\gamma$-ray emission
in the NPS region and the shape of this excess is similar to those seen in
synchrotron emission \citep{2009arXiv0912.3478C}. This may buttress the 
reheating picture of Loop I. We set $2\times10^5$ years as the age of NPS which 
is suggested by \citet{egger95}, and the distance to NPS is 100 pc. Like the 
previous model, we set $\gamma_{\rm loop}$, $W_{\rm loop}$, and $E_{c, {\rm 
loop}}$ as free parameters.

As presented in Fig. \ref{fig:loop1}, the total anisotropy of this model is 
clearly under the constraint of Fermi-LAT. Below 1 TeV, the anisotropy of this 
model is dominated by the background SNRs. The contribution of Loop I (NPS) 
to the total anisotropy is smaller than the background component, because of 
the old age and close distance of Loop I (NPS). This indicates that the 
anisotropy of the background SNRs is indeed non-negligible. As shown in the top 
right of Fig. \ref{fig:loop1}, the total anisotropy below 100 GeV is possible 
to be detected by HERD with 3 years of observation. At the same time, the 
angular distribution of intensity at 100 GeV (bottom of Fig. \ref{fig:loop1}) 
is distinct from that of the \textit{Monogem Ring} scenario, since Monogem Ring 
locates in almost the opposite direction to the Galactic center. Thus the 
future measurement of HERD may distinguish between this model and the 
\textit{Monogem Ring} model.  

\subsection{TeV Region}
\label{subsec:TeV}
The $e^-+e^+$ spectrum above TeV has been measured by several experiments, 
while there is still inconsistency among these measurements. DAMPE has 
confirmed a clear spectral break in $\sim$1 TeV \citep{nature}, which is in 
agreement with the earlier result of H.E.S.S \citep{hess08}. Thus we add 
several TeV sources respectively on a background with a break power-law form to 
discuss the possible spectral features in the TeV region. For the background, 
the spectral index below 1 TeV is assumed to be $-3.17$ as measured by AMS-02, 
and soften by 1 above 1 TeV. The anisotropy is assumed to be dominated by the 
TeV local sources.

\begin{figure*}
\centering
\includegraphics[width=0.4\textwidth]{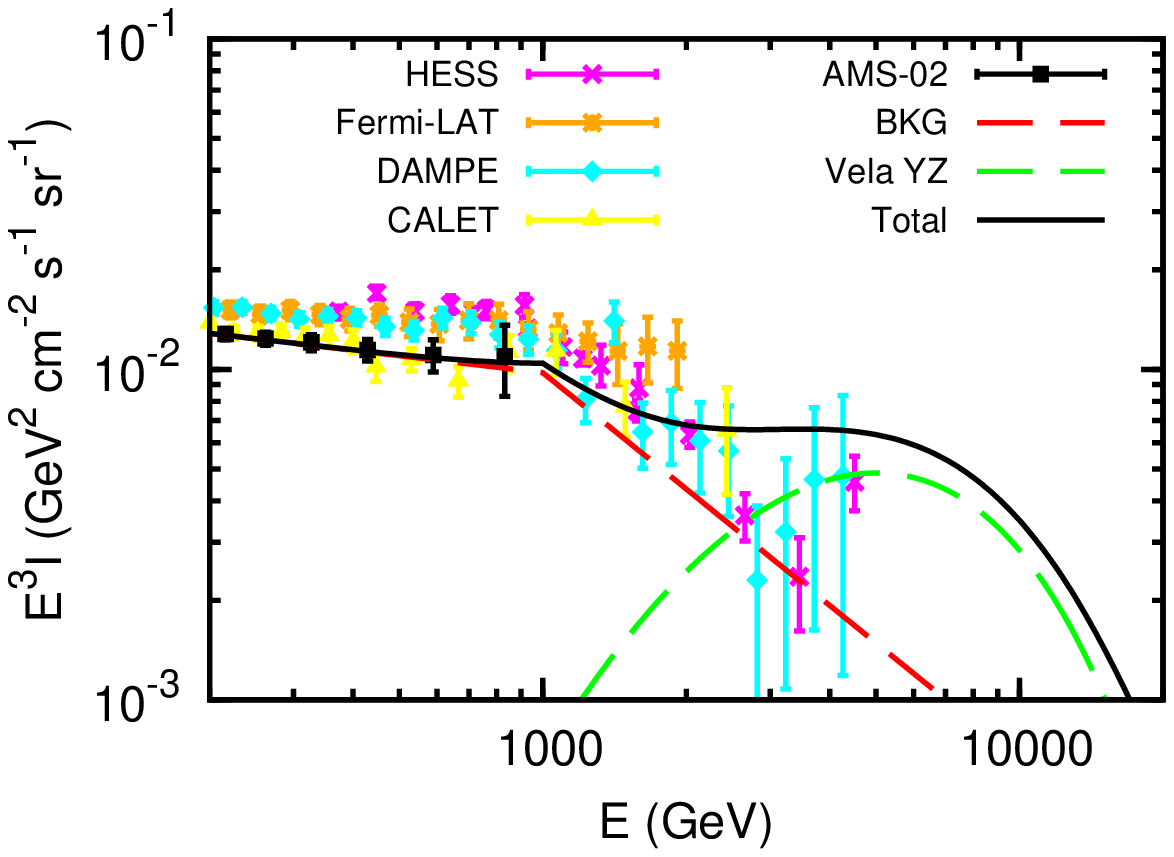}
\includegraphics[width=0.4\textwidth]{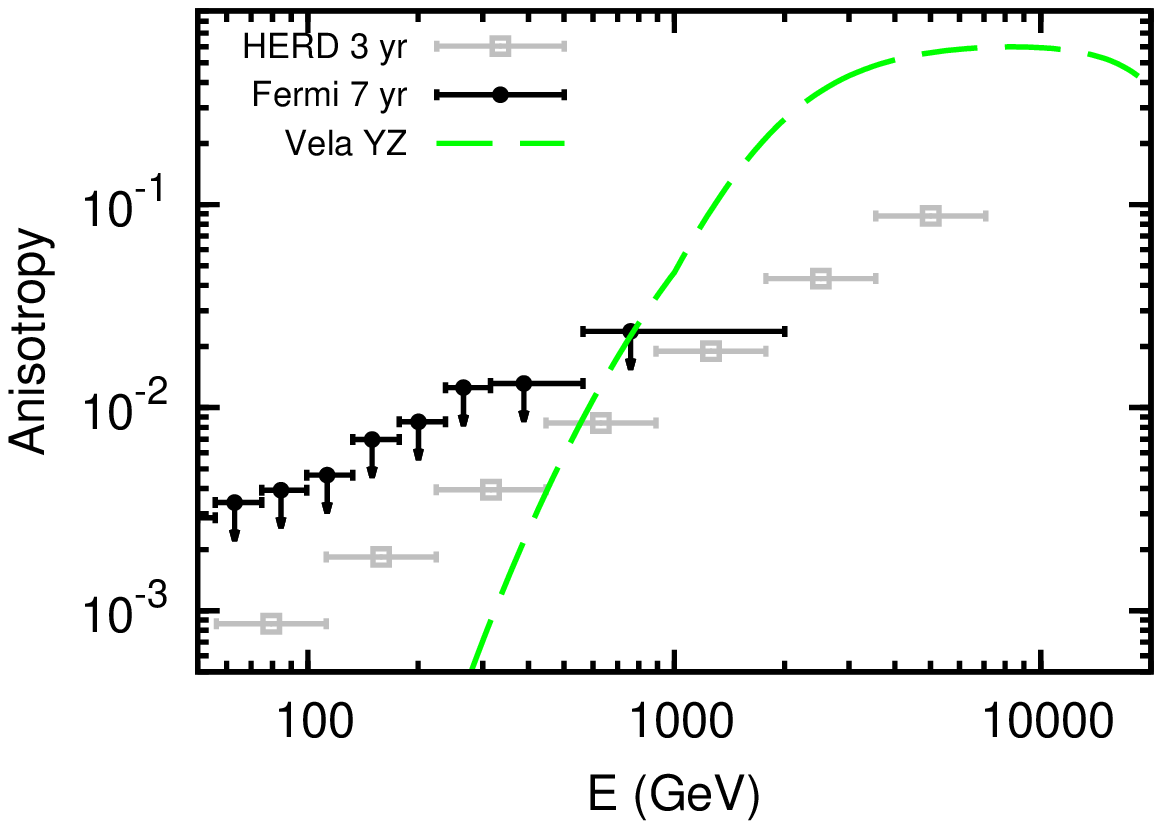}
\caption{The scenario that Vela YZ produces TeV spectral feature. A 
considerable injection delay of 10 kyr is assumed. Note the background 
component is a break power-law spectrum, which has different meaning with that 
in the figures of Section \ref{subsec:subteV}.}
\label{fig:vela_TeV}
\end{figure*}

\subsubsection{Vela YZ}
\citet{koba04} have shown the scenario that if we take into account the release 
time, that is, the time delay of electron injection, Vela YZ can contribute 
significantly in the TeV region. Particles may begin to escape from the shock 
front of SNR when the velocity of the shock has dropped to the order of the 
Alfv$\rm\acute{e}$n velocity of the ISM \citep{dorfi00}. The mean 
Alfv$\rm\acute{e}$n velocity of ISM can be calculated by $v_A=2.18\times10^5$ 
cm s$^{-1}$ $(m_i/m_p)^{-1/2}\,(n_{\rm ISM}/{\rm cm^{-3}})^{-1/2}\,(B/\mu{\rm 
G})$, where $m_i$ is the ion mass, $m_p$ is the mass of proton, $n_{\rm ISM}$ 
and $B_{\rm ISM}$ are number density and magnetic field of ISM respectively. 
The dynamics of expansion of Vela SNR suggests $n_{\rm ISM}\leq0.01$ cm$^{-3}$ 
\citep{sushch11}, and if we assume $B_{\rm ISM}$ to be 10 $\mu$G, the 
Alfv$\rm\acute{e}$n velocity of the surrounding medium should be 
$\sim1\times10^7$ cm s$^{-1}$. The shock velocity of Vela YZ region is observed 
to be $6\times10^7$ cm s$^{-1}$ \citep{sushch11}, and the shock velocity 
evolves 
with $t^{-3/5}$ in Sedov phase. So the initial velocity should be large than 
$10^8$ cm s$^{-1}$, which is much faster than the Alfv$\rm\acute{e}$n velocity 
of ISM. This indicates that a considerable release time for Vela YZ is 
reasonable.

In this model, we assume a release time of 10 kyr for Vela YZ, which 
corresponds to a young injection age of 1 kyr. We take a typical value of 
2.0 for the injection spectral index of Vela YZ, and the cut-off energy is 
assumed to be 4 TeV as we explained in the previous subsection. Fig. 
\ref{fig:vela_TeV} presents the $e^-+e^+$ spectrum and the anisotropy of this 
model. The TeV spectrum has a bump feature compared with a power-law form, 
although Vela YZ does not stand out enough from the background. If the 
injection age of Vela YZ is larger, its flux of lower energy electrons will 
increase. Then Vela YZ will be mixed with the background and no remarkable 
feature is expected. At the same time, the anisotropy of this model conflicts 
with the upper limit of the last energy bin of Fermi-LAT. The reason is 
indicated by equation (\ref{eq:ani_t}): the $\Delta_i$ of Vela YZ with such a 
young injection age is very large ($\sim0.7$), and the flux of Vela YZ is 
non-neglectable in sub-TeV.

\begin{figure*}
\centering
\includegraphics[width=0.4\textwidth]{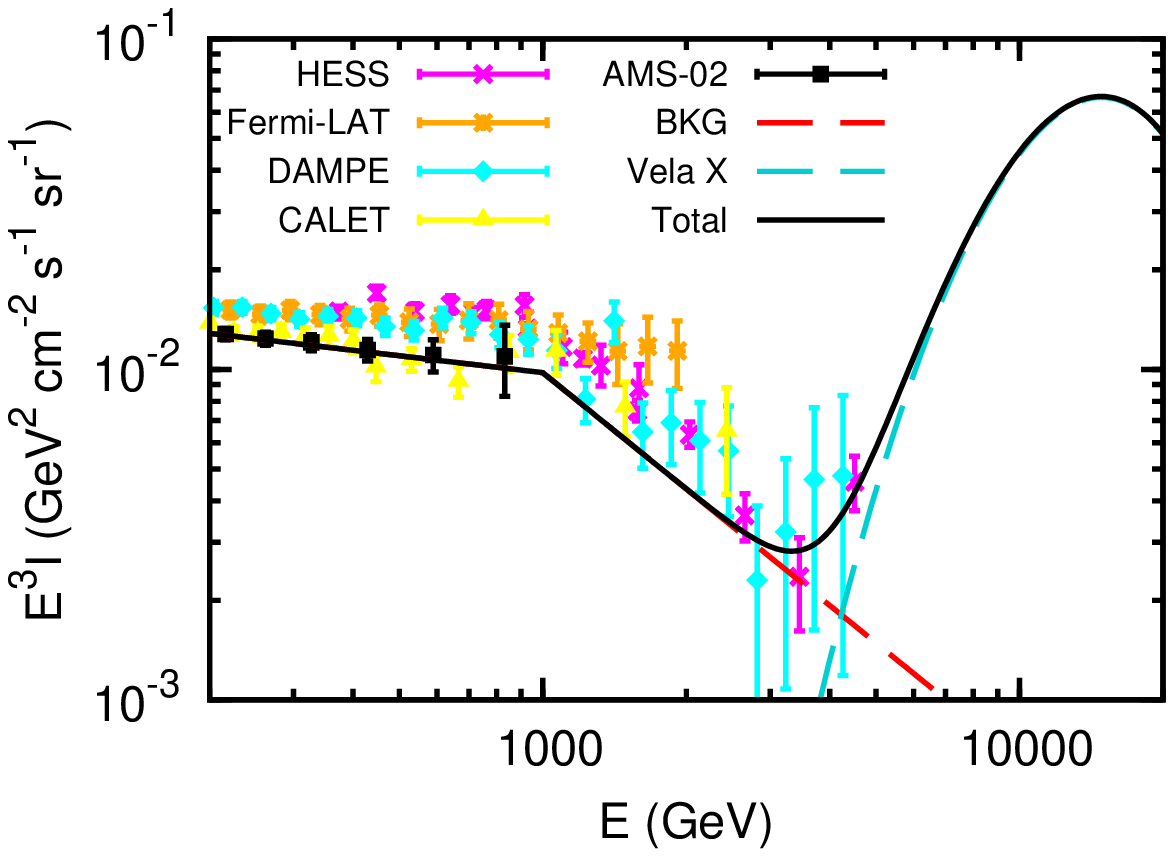}
\includegraphics[width=0.4\textwidth]{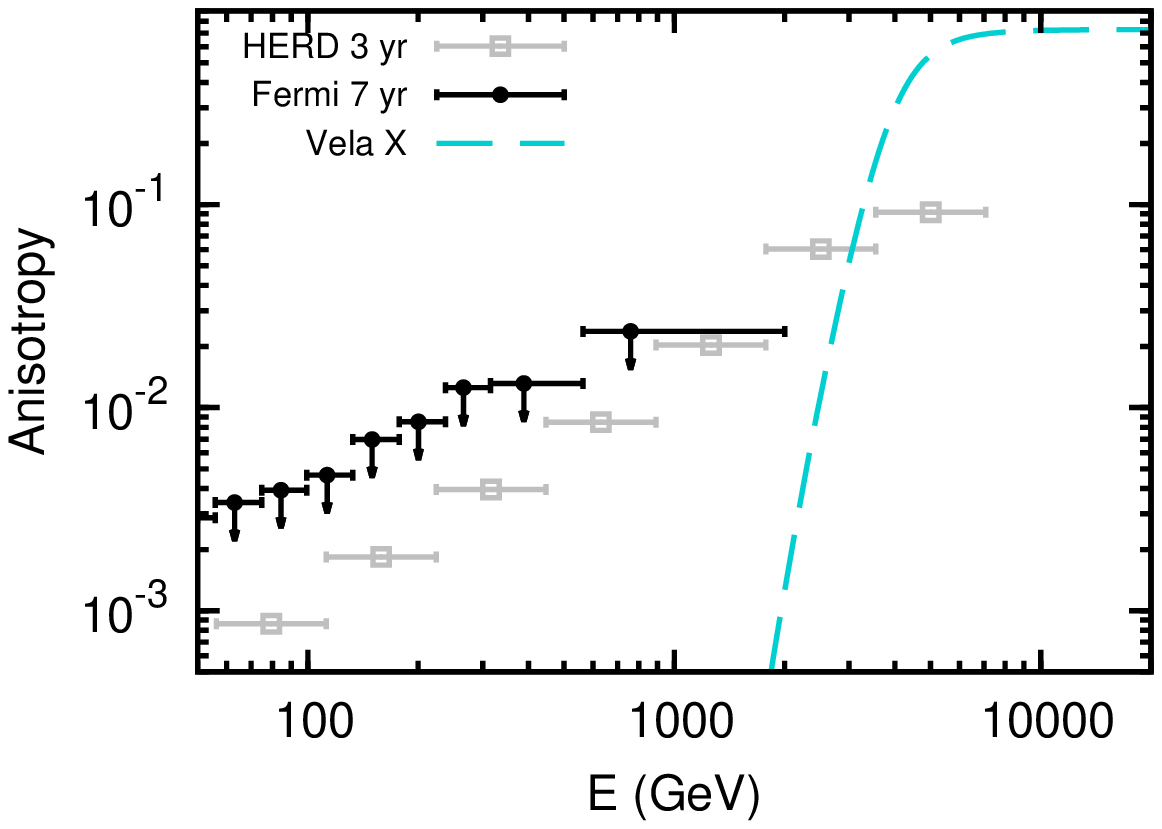}
\includegraphics[width=0.4\textwidth]{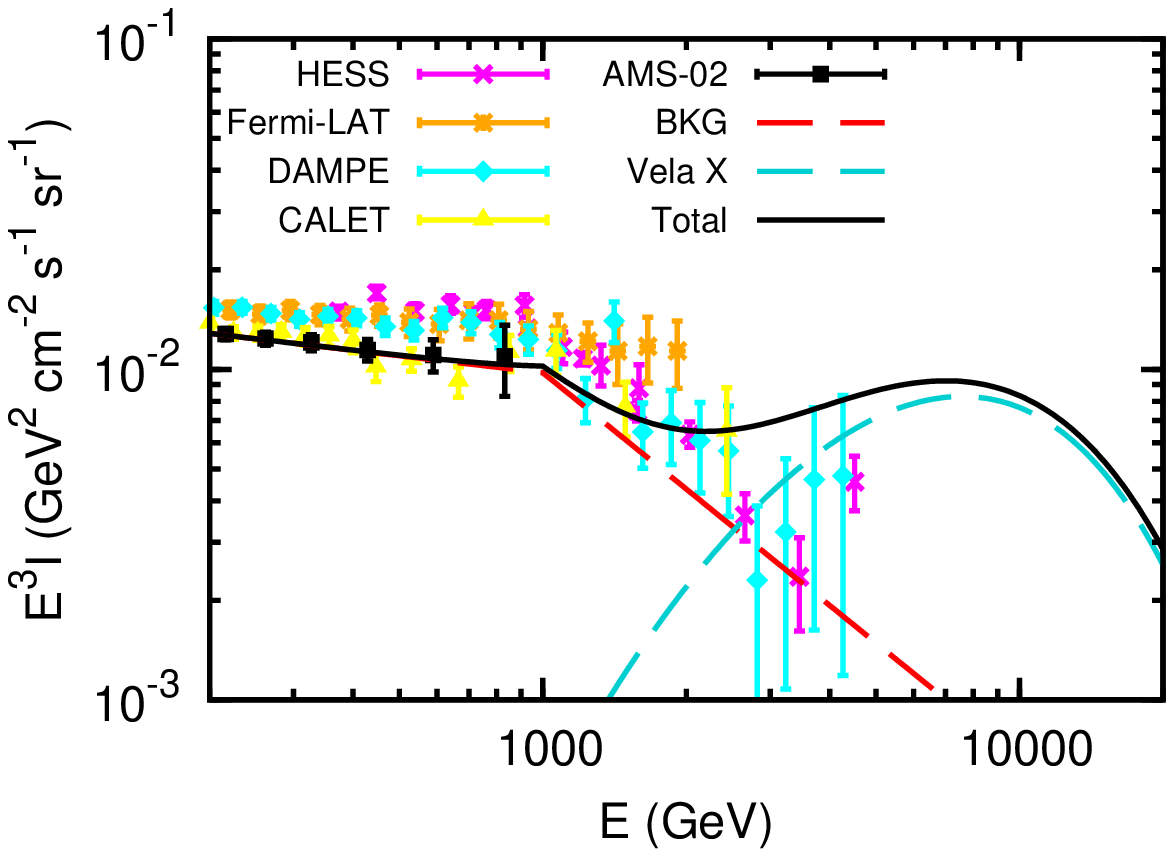}
\includegraphics[width=0.4\textwidth]{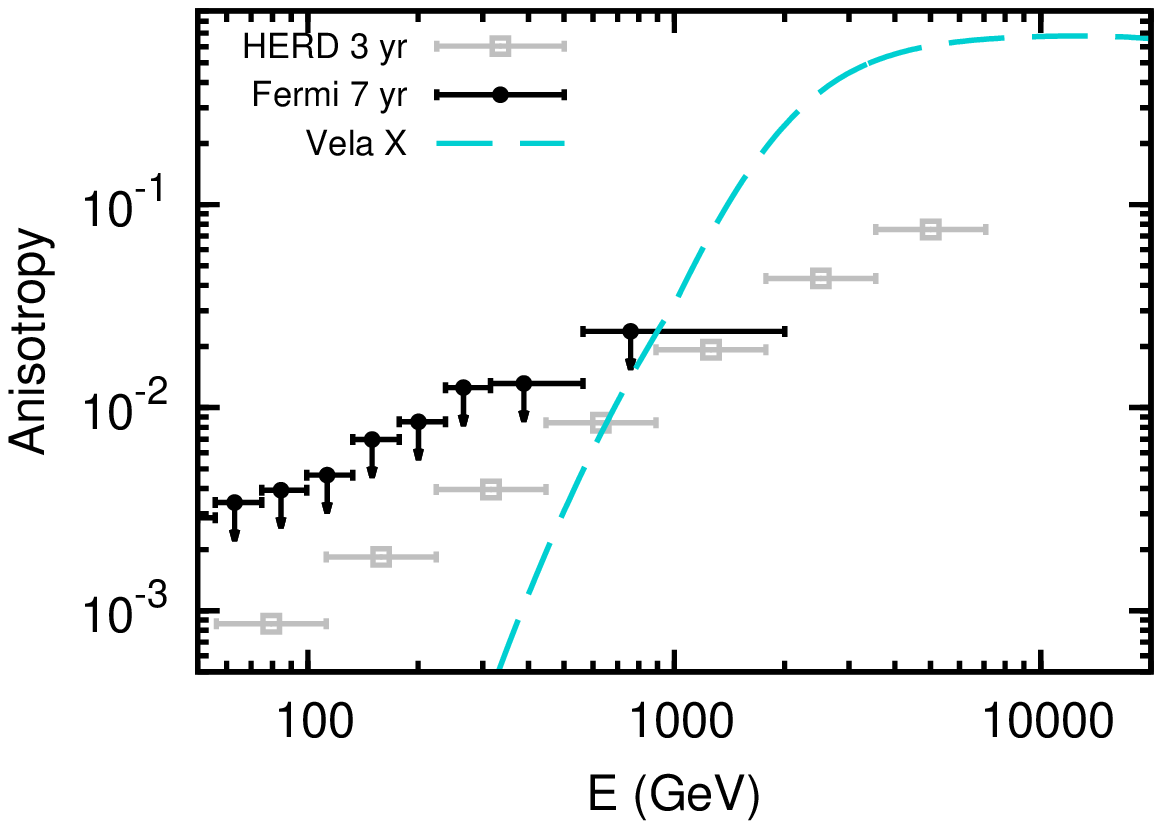}
\caption{The scenario that Vela X produces TeV spectral feature. Top graphs: 
the $e^-+e^+$ spectrum and anisotropy spectrum using the model of 
\citet{hinton11}, and release time of 7 kyr is adopted; bottom graphs: the 
same with the top ones, but we use the diffusion coefficient of our own, and 
assume a leptonic injection energy of $10^{47}$ erg and a release time of 10 
kyr.}
\label{fig:velax_TeV}
\end{figure*}

\subsubsection{Vela X}
Vela X consists of two component: a halo and a 'cocoon' 
\citep{jager08,hinton11}. However, the low cut-off energy of 100 GeV for the 
former and the small total leptonic energy of $10^{46}$ erg for the later 
prevent Vela X to create a distinctive spectral feature in TeV region. 
\citet{hinton11} point out that the low cut-off energy for the halo component 
may be attributed to an energy dependent escape which happens at the time of 
the crush of the original PWN. They assume a spectral index of 1.8,
a cut-off energy of 6 TeV, and a total leptonic energy of $6.8\times10^{48}$ erg
in their model. After considering the release time of electrons, their result 
shows that Vela X can produce a prominent TeV spectrum of $e^-+e^+$.
Nevertheless, another precondition of their spectral shape is the small
diffusion coefficient they adopt, which is an order of magnitude smaller than
ours. We have explained in F17 that if we use our diffusion coefficient
with their injection spectral parameters, the predicted $e^-+e^+$ flux of Vela
X is too large even in sub-TeV, and has a serious contradiction with the AMS-02 
data. We should also point out that the total leptonic energy of 
$6.8\times10^{48}$ erg may be a too large value, since the spin-down luminosity
of Vela X given by the ATNF catalog is $6.92\times10^{36}$ erg, corresponding
to a spin-down energy of merely $5\times10^{48}$ erg. Besides, the four years 
of observation of Fermi-LAT and the radio data of Vela X derive a total 
leptonic energy of $9\times10^{47}$ erg \citep{grondin13}. This may be regarded 
as the upper limit of the leptonic injection energy.

Here we present two scenarios based on different treatments to Vela X: the one
is the model given by \citet{hinton11} with an injection age of 3 kyr; in the 
other one, we adopt the diffusion coefficient of our own. For the later, a 
small leptonic injection energy is indispensable to avoid the conflict with the 
present $e^-+e^+$ measurements. We set an injection energy of $10^{47}$ erg and 
a smaller injection age of 1 kyr for the second model. The predicted $e^-+e^+$ 
spectra and anisotropies are shown in Fig. \ref{fig:velax_TeV}. For the first 
model, Vela X produces a prominent enough feature in the $e^-+e^+$ spectrum, 
and the predicted anisotropy is not constrained by the upper limits of 
Fermi-LAT. The steep spectrum of Vela X should be ascribed to the much smaller 
diffusion coefficient. Alternatively, a very hard injection spectral index 
can also lead to a steep spectral feature. The second model in Fig. 
\ref{fig:velax_TeV} is similar with the previous model of Vela YZ because of 
the 
similarity of their parameters.

\begin{figure*}
\centering
\includegraphics[width=0.4\textwidth]{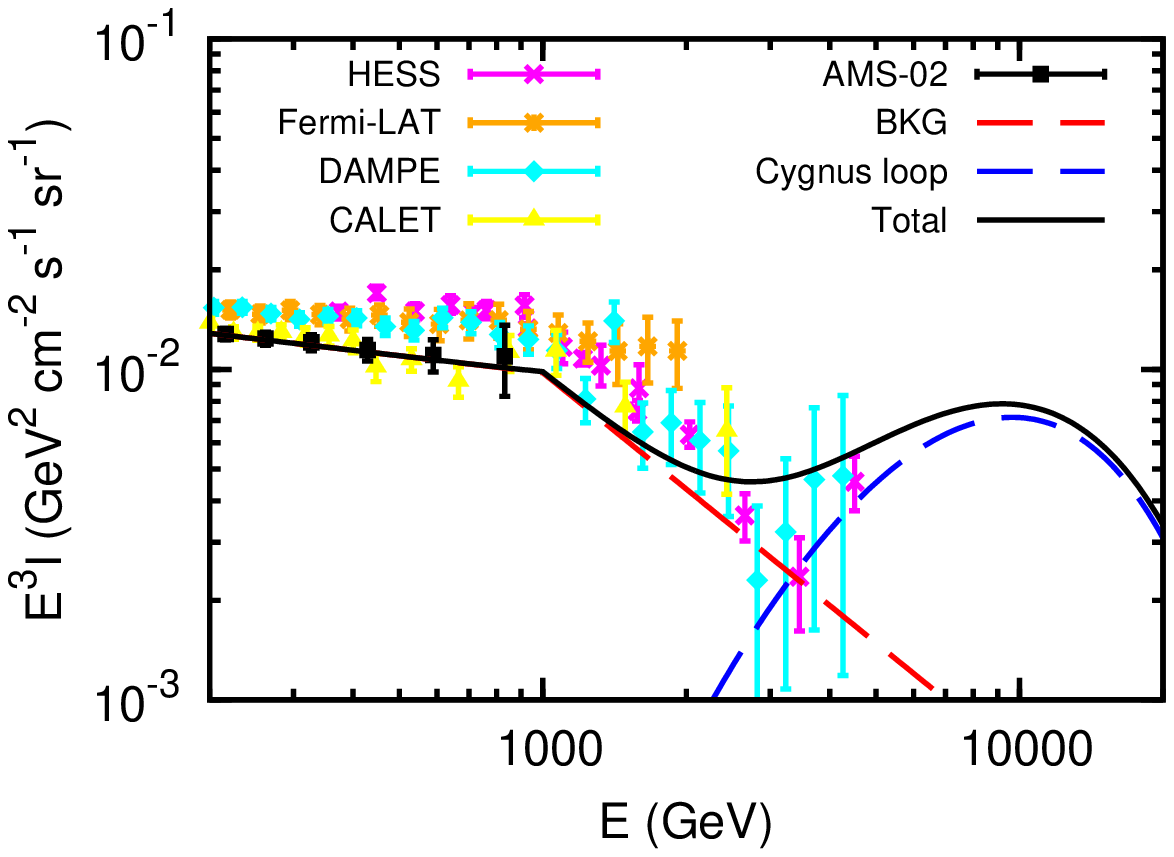}
\includegraphics[width=0.4\textwidth]{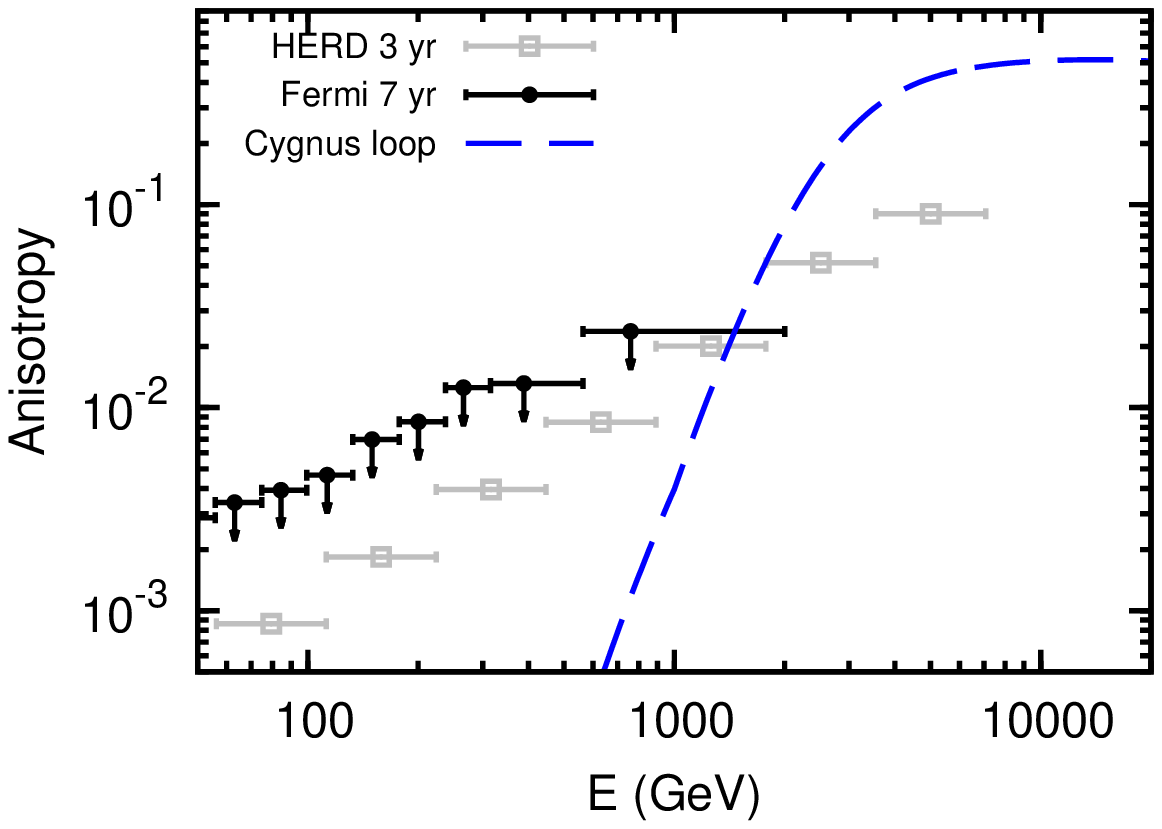}
\caption{Cygnus Loop overlays on \textit{Loop I (NPS)} model. The
cut-off energy of Cygnus Loop is 8 TeV, other injection parameters are kept as
the fitted value in F17: $\gamma=1.99$, $Q_0=10^{50}$ GeV$^{-1}$. Left: the
$e^-+e^+$ spectrum compared with experimental data; right: the corresponding
anisotropy, upper limits of Fermi-LAT, and expected detection ability of HERD.}
\label{fig:cygnus_TeV}
\end{figure*}

\subsubsection{Cygnus Loop}
Another famous nearby SNR---Cygnus Loop---does not appear in the preceding part 
of the text due to its very low cut-off energy (72 GeV) given by the multi-band 
fitting in F17. However, there is no available X-ray spectrum of the global 
region of Cygnus Loop, so the fitted cut-off energy may not be so compelling. 
If we calculate the cut-off energy of Cygnus Loop with evolution model of SNR 
\citep{yamazaki06}, it should be in the TeV region. Then Cygnus Loop may become 
a prominent TeV electron contributor. In this model, we keep the parameters of 
Cygnus Loop fitted in F17, except for the cut-off energy. The size of Cygnus 
Loop is approximately 200 arcmin, then a distance of 540 pc corresponds to a 
radius of 15 pc. The velocity of shock wave can be estimated by $0.4\,R/t$ 
\citep{sushch11}, where $R$ is the radius of the shell and $t$ is the age the 
SNR. We derive a shock velocity of $6\times10^7$ cm s$^{-1}$, and the magnetic 
field given by the multi-band fitting is 9.7 $\mu$G, so the cut-off energy is 
estimated to be approximately 8 TeV. We set a release time of 7 kyr, 
corresponding to an injection age of 3 kyr. The $e^-+e^+$ spectrum and 
corresponding anisotropy are presented in Fig. \ref{fig:cygnus_TeV}. The 
spectral feature generated by Cygnus Loop is prominent, since the distance of 
Cygnus Loop is larger compared with Vela and so its spectrum can stand out from 
the background. We also find the anisotropy of this model only conflicts 
slightly with the upper limits of Fermi-LAT, compared with the model of Vela YZ 
and the second scenario of Vela X.

From all the models above, we obtain the constraint on possible TeV 
spectral features with the restriction of Fermi-LAT: the potential feature 
should be steep like the first case of Vela X, or it should appear in high 
energy region of $E>5$ TeV. The corresponding physical conditions are: a very 
small diffusion coefficient, or a very hard injection spectral index, or a 
relatively far distance of the source. Another obvious result is that the 
anisotropies of all these TeV models can be detected by HERD. If the anisotropy 
is detected along with a spectral feature in the future, we may even give 
constraints on the astrophysical models. For example, assuming a prominent 
feature is detected in the TeV region, and the $\bm{n}_{\rm max}$ of the 
intensity distribution points at the location of Cygnus Loop in the 
corresponding energies. This would indicates that Cygnus Loop should be the 
source accounting for this spectral feature. Then as we described above, the 
electron release time of Cygnus Loop can be derived from the spectral shape, 
which is a parameter that can hardly be determined by electromagnetic 
observations. 

\begin{table*}
\centering
 \begin{tabular}{cccccccc}
  \hline
  Name & $l$($^\circ$) & $b$($^\circ$) & $r$(kpc) & $t$(kyr) & $W_{\rm p}$
($10^{49}$ erg) & $\gamma_{\rm pwn}$ & $\eta_{\rm pwn}$ \\
  \hline
  J0940-5428 & 277.5 &  $-1.3$ & 0.38 & 42 & 1.34 & 1.5 & 0.036 \\
  \hline
  Geminga & 195.1 & $+4.3$ & 0.25 & 342 & 1.23 & 2.0 & 0.88 \\
  \hline
  B1001-47 & 276.0 & $+6.1$ & 0.37 & 220 & 0.480 & 2.0 & 0.78 \\
  \hline
  J2043+2740 & 70.6 & $-9.2$ & 1.48 & 1200 & 25.8 & 2.0 & 0.38 \\
  \hline
 \end{tabular}
 \caption{Members of the Multi-PWN model in Section \ref{subsec:PWNe}. Their
position, distance, and age are referred to the ATNF catalog. The last two
columns are given by the fitting described in Section \ref{subsec:PWNe} (the
upper and lower bounds of $\gamma_{\rm PWN}$ are set to be 2.0 and 1.5 in the
fitting process).}
\label{tab:PWNe}
\end{table*}

\section{Discussion}
\label{sec:discussion}

\subsection{Multi-PWN Model}
\label{subsec:PWNe}
Although the \textit{Monogem Ring} model and the \textit{Loop I (NPS)} model 
can explain the AMS-02 data and their anisotropies are entirely under the 
upper limits of Fermi-LAT, they predict a too steep spectral cut at 1 TeV 
compared with all those TeV measurements. The reason is that all the local
sources in sub-TeV---MR, Loop I, and Geminga---begin to descend just below 1 TeV
in the spectrum. Geminga has the sharpest decline due to its relatively old age.
Here we test a model that consists of a group of PWNe, instead of the case of a 
single PWN applied above. Young member(s) of the PWN group may help to 
contribute to the flux around 1 TeV.

We divide the energy range from 10 GeV to 1 TeV into four bins with equal 
length in logarithmic scale. For each bin, we calculate the integrated positron 
flux for all the PWNe in our sample, with a uniform spectral index of 1.8. Then 
all the PWNe are ranked by their integrated flux in each bin. We sum their rank 
of the four bins for each PWN individually, and the ten with the smallest 
summed rank are selected as our candidates. They are: J0940-5428,
B1055-52, J0633$+$1746 (Geminga), B0355$+$54, B1001-47, B0656$+$14
(Monogem), J0538$+$2817, J1732-3131, J2043$+$2740, B1742-30. 

\begin{figure*}
\centering
\includegraphics[width=0.4\textwidth]{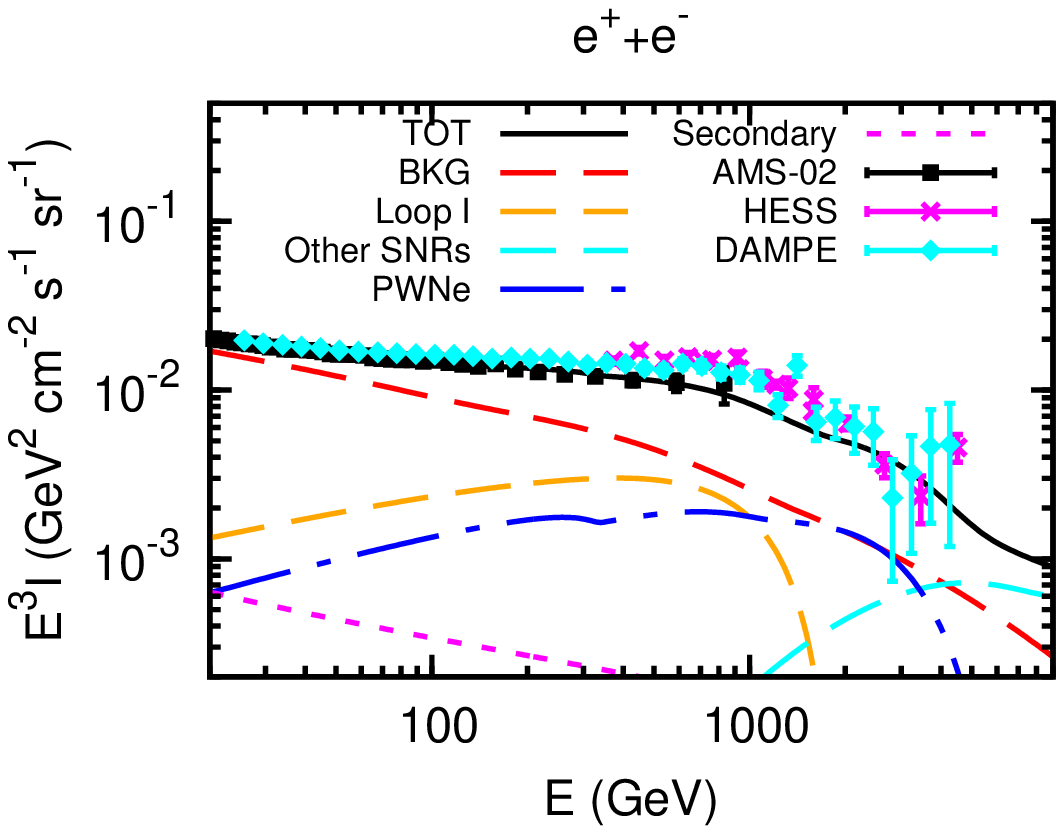}
\includegraphics[width=0.4\textwidth]{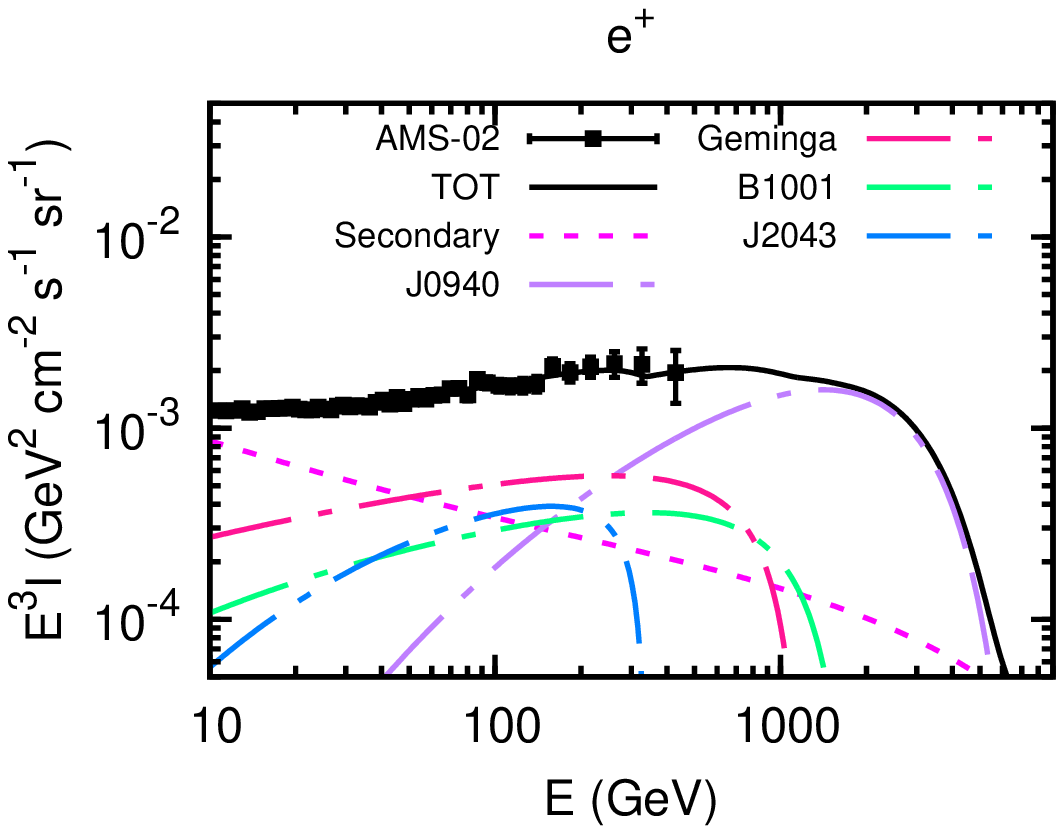}
\includegraphics[width=0.4\textwidth]{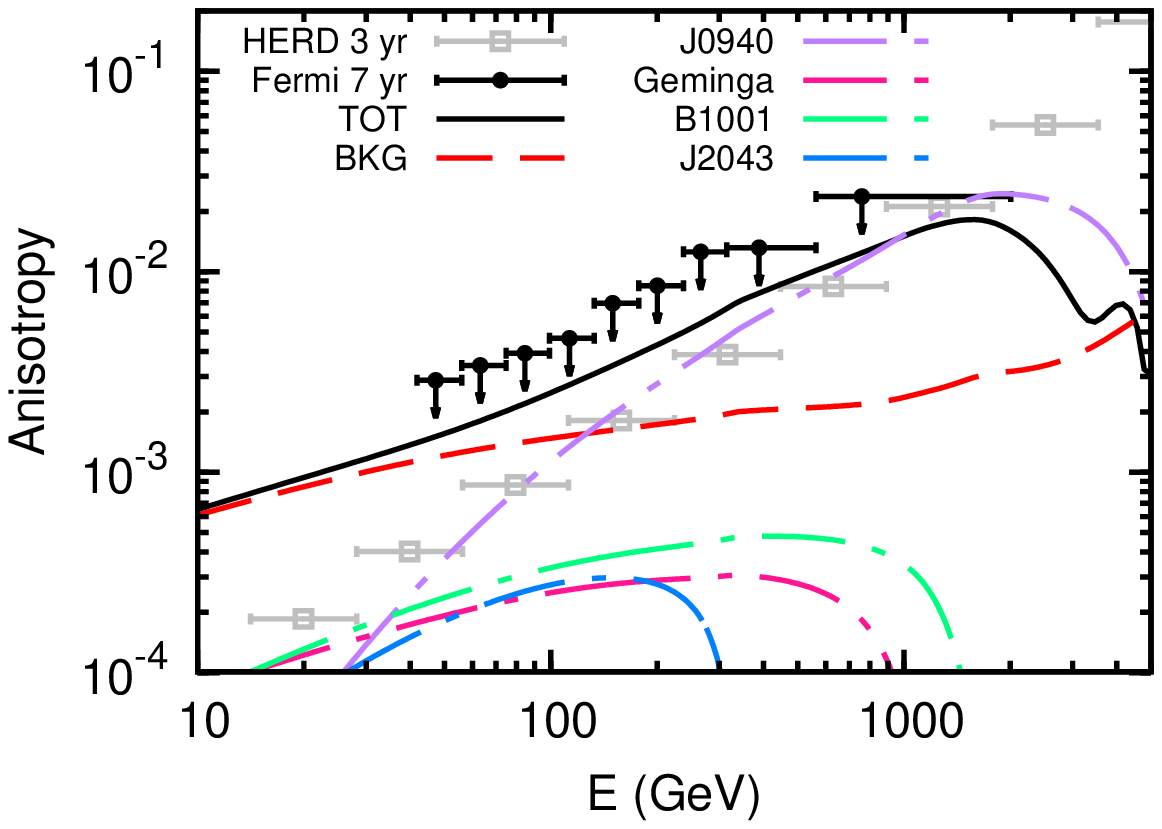}
\caption{Same as the \textit{Loop (NPS)} model in Section \ref{subsec:subteV}, 
but replace the single PWN with a group of PWNe. The positron spectrum of each 
PWN is shown in the top right.}
\label{fig:pow_pwn}
\end{figure*}

We replace Geminga with the ten PWNe in the \textit{Loop I (NPS)} model, to fit 
the AMS-02 data. The spectral index and conversion efficiency of each PWNe are 
set to be free, and the cut-off energy is fixed at 2 TeV for all the PWN 
members. Other parameters in the model remain unchanged as given in Table 
\ref{tab:sub-TeV}. In the new fitting procedure, the $\eta_{\rm pwn}$ of six 
PWNe converge to zero, which indicates the other four sources are enough to 
explain the AMS-02 data. We present the fitting result along with the 
information of these four PWNe in Table \ref{tab:PWNe}. The leptonic spectra of 
the multi-PWN model and the anisotropy of $e^-+e^+$ are shown in Figure 
\ref{fig:pow_pwn}. We find that the spectral break in 1 TeV is milder compared 
with the original \textit{Loop I (NPS)} model, and is in a better consistency 
with current measurements. Besides, the $e^-+e^+$ anisotropy of this multi-PWN 
model is also under all the
upper limits of Fermi-LAT. 

The flux increase in $\sim$ 1 TeV owes much to the source J0940-5428 which has a
significant contribution up to several TeV. J0940-5428 is a Vela-like pulsar 
due to its fast spinning, relatively young age, and large spin-down luminosity 
\citep{2007AJ....134.1231C}. The ATNF catalog now provides a much closer 
distance of 0.38 kpc for J0940-5428, compared with the old estimation of $\sim$ 
4 kpc. This is crucial to update the status of J0940-5428 in the $e^\pm$ 
spectrum. However, we still have much less knowledge of J0940-5428 at present 
than well-studied sources like Geminga. Researches have shown that J0940-5428 
may not be surrounded by an observable synchrotron nebula as the case of Vela X 
or Geminga \citep{2014ApJ...793...89W}.

\subsection{The Case of Featureless TeV Spectrum}
Since the contributors of TeV $e^\pm$ are few, remarkable spectral structures 
are often expected. However, the latest preliminary $e^-+e^+$ spectrum of 
H.E.S.S indicates no prominent feature up to $\sim20$ TeV \citep{hess17}. 
If the TeV $e^-+e^+$ spectrum is indeed proved to be  featureless in the 
future, we can hardly discriminate local sources by the $e^-+e^+$ spectrum. 
In this case, the measurement of anisotropy should be the most important tool 
of investigating the origin of TeV $e^\pm$. Here we discuss the possibility of 
detecting the anisotropy for a $e^-+e^+$ spectrum with no prominent feature.

\begin{figure*}
\centering
\includegraphics[width=0.4\textwidth]{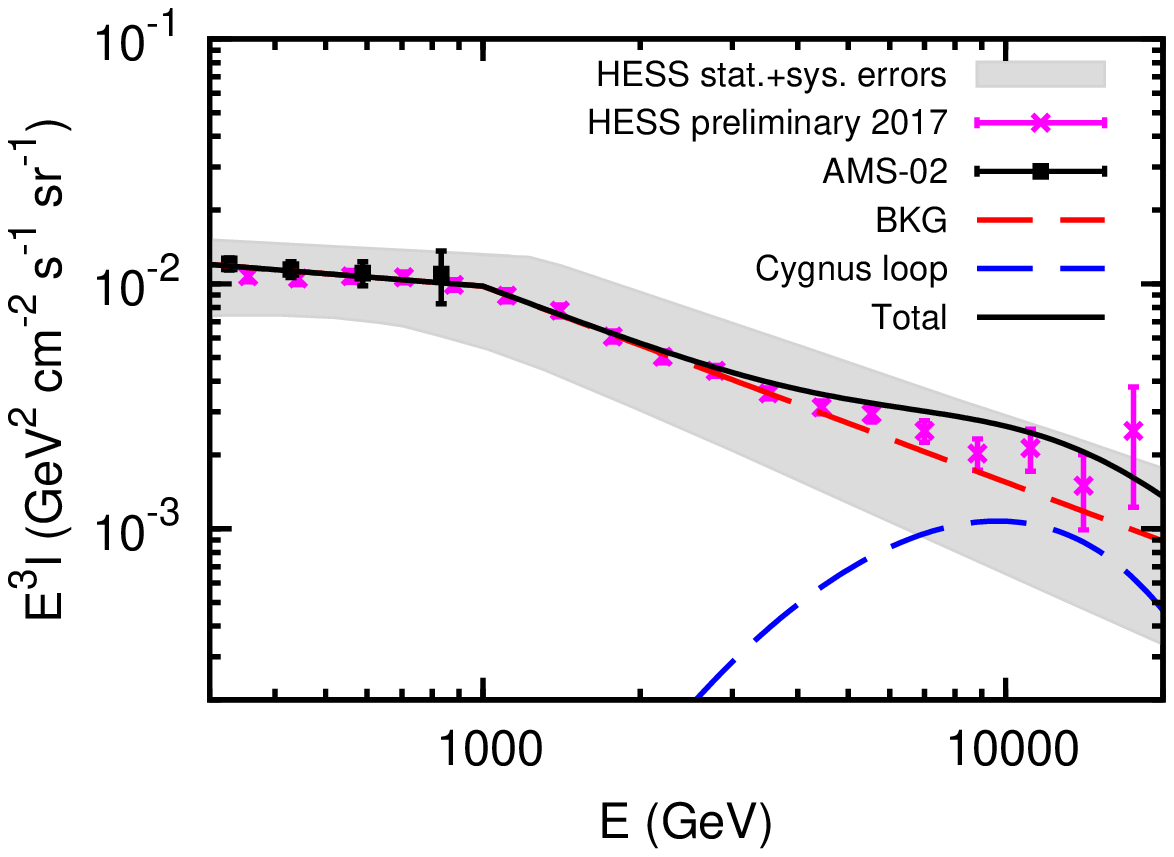}
\includegraphics[width=0.4\textwidth]{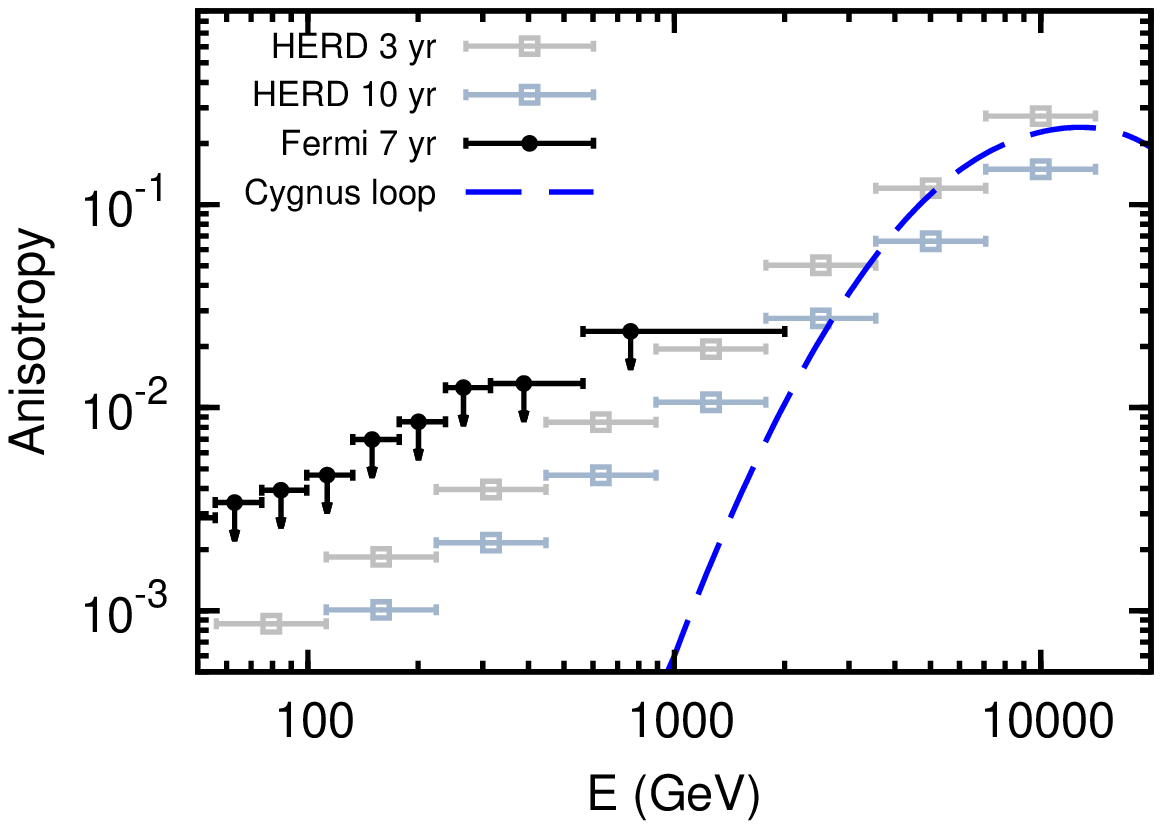}
\caption{An example with no remarkable feature in the TeV $e^-+e^+$ spectrum. 
The background of this model is a break power-law spectrum with a spectral 
softening from $-3.17$ to $-3.8$ at 1 TeV.}
\label{fig:hess}
\end{figure*}

As an example, we add Cygnus Loop on a break power-law background to 
accommodate the preliminary spectrum of H.E.S.S. The spectral index of the 
background is set to be $-3.17$ and $-3.8$ in the region below and above 1 TeV, 
respectively. The normalization of Cygnus Loop is assumed to be 15\% of the 
case in Section \ref{subsec:TeV}, to ensure no prominent feature in the total 
spectrum. The anisotropy is still assumed to be dominated by Cygnus Loop. The 
spectrum and anisotropy of this case are shown in Fig. \ref{fig:hess}. As 
can be seen, the slight spectral bump produced by Cygnus Loop is within the 
error band of H.E.S.S. At the same time, the anisotropy is hopeful to be 
detected by HERD above $\sim5$ TeV.

\section{Conclusion}
\label{sec:conclusion}
In this paper, we first use the latest anisotropy result of Fermi-LAT to test 
local source models aimed to explain the leptonic data of AMS-02. Our results 
show that the anisotropy spectrum of the \textit{Vela YZ} model reaches the 
exclusion limit at 95\% C.L. given by Fermi-LAT, which means the result of 
Fermi-LAT disfavors Vela YZ as the dominant local SNR in the sub-TeV region. 
The other two models, where 
Monogem Ring or Loop I (NPS) plays the role of the dominant local SNR, remain 
safe under the restriction of the Fermi-LAT result. This implies the extra 
electron excess in sub-TeV should be explained by a relatively old source. The 
next generation instrument HERD is expected to have a better sensitivity than 
Fermi-LAT. We then estimate the capability of anisotropy detection of HERD, and 
find HERD is sensitive enough to discriminate between the \textit{Monogem Ring} 
model 
and the \textit{Loop I (NPS)} model, as the predicted angular intensity maps of 
these two scenarios are remarkably different. We also point out that the SNR 
background has a considerable contribution to the anisotropy which is even 
larger than that of some discrete local sources like Geminga, so this component 
should not be neglected in the calculation, especially in sub-TeV.

Since fewer local sources can contribute significantly to the TeV $e^-+e^+$ 
spectrum, spectral features are expected in the TeV region. We discuss several 
cases of remarkable features in TeV by adding dominant local sources on a 
break power-law background. The predicted anisotropies of some cases conflict 
with the upper limits of Fermi-LAT, and we find the conditions for a remarkable 
TeV feature which satisfies the anisotropy constraint by Fermi-LAT: if Vela YZ 
or Vela X is the dominant TeV source, a considerable electron injection delay 
(10 kyr) is necessary, and a much smaller diffusion coefficient than usual or a 
very hard injection spectrum is needed to avoid the anisotropy constraint; a 
dominant source with relatively farther distance is also a solution, like 
Cygnus Loop, which can produce a spectral feature in higher energies ($>$ 5 
TeV) 
considering an injection delay. Moreover, the predicted anisotropies of all 
those TeV models can be detected by HERD. So if we could detect remarkable 
spectral features along with the anisotropy in the future, we may even give 
constraints on some physical parameters, such as the injection delay of the 
source, or the diffusion coefficient of high energy $e^\pm$.

Besides, the latest preliminary $e^-+e^+$ spectrum of H.E.S.S indicates no 
prominent feature in the TeV region. We discuss a case with no significant 
structure in the $e^-+e^+$ spectrum, and find that HERD will likely detect 
the anisotropy of a local source in this case. Thus, the possibility of 
ascertaining the local source is still retained even in the case of a 
featureless $e^-+e^+$ spectrum, and HERD may play an important role in the 
study of the origin of high energy CR $e^\pm$.

\section*{Acknowledgement}
We thank Ming Xu for providing the simulated performance of HERD. This work is 
supported by the National Key Program for Research and Development 
(No.~2016YFA0400200) and by the National Natural Science Foundation of China 
under Grants No.~U1738209,~11475189,~11475191, and is supported in part by the 
CAS Center for Excellence in Particle Physics (CCEPP).

\end{document}